\documentclass[reprint, aps, pra, twoside, floatfix, nofootinbib, longbibliography, superscriptaddress]{revtex4-2}
\usepackage{amsmath, amssymb, amsbsy}
\usepackage{graphicx}
\usepackage[usenames, dvipsnames]{color}
\usepackage{mathtools}
\usepackage{orcidlink}
\usepackage{xcolor}
\usepackage{bbold}
\usepackage{dcolumn}
\usepackage{hyperref}
\hypersetup{colorlinks=true, linkcolor = blue, citecolor = blue, urlcolor = blue}
\newcommand{\Tr}{\mathrm{Tr}}		
\newcommand{\mathhc}{\textrm{H.c.}}	
\newcommand{\dd}{\mathrm{d}}		
\newcommand{\ee}{\mathrm{e}}		
\newcommand{\ii}{\mathrm{i}}		
\newcommand{\opH}{\hat{H}}			
\newcommand{\opD}{\mathcal{D}}		
\newcommand{\Phib}{\Phi_\mathrm{b}}	
\newcommand{\res}{\mathrm{r}}  		
\newcommand{\osc}{\mathrm{o}}       
\newcommand{\pd}{\mathrm{d}}		
\newcommand{\cpl}{\mathrm{c}}		
\newcommand{\JJ}{\mathrm{J}}		
\newcommand{\zpf}{\mathrm{zpf}}		
\newcommand{\eff}{\mathrm{eff}} 	
\newcommand{\clk}{\mathrm{cl}}      
\newcommand{\phibtm}{\phi_\cup}     
\newcommand{\phitop}{\phi_\cap}     
\newcommand{\cA}{\mathcal{A}}
\newcommand{\cE}{\mathcal{E}}
\newcommand{\cG}{\mathcal{G}}
\newcommand{\cH}{\mathcal{H}}
\usepackage[normalem]{ulem}
\begin{document}
\title{Detector of microwave photon pairs based on a Josephson photomultiplier}
\author{E.~V.~Stolyarov\,\orcidlink{0000-0001-7531-5953}}
\email{eugene.stolyarov@bitp.kyiv.ua}
\affiliation{Quantum Optics and Quantum Information Group, Bogolyubov Institute for Theoretical Physics, National Academy of Sciences of Ukraine, vul. Metrolohichna 14-b, Kyiv 03143, Ukraine}
\author{R.~A.~Baskov\,\orcidlink{0000-0001-9285-9469}}
\email{roman.baskov@yale.edu}
\affiliation{Department of Physics and Applied Physics, Yale University, P.O. Box 208120, New Haven, Connecticut 06520-8120, USA}
\affiliation{Yale Quantum Institute, 17 Hillhouse Ave., PO Box 208334, New Haven, CT 06520-8263 USA}
\begin{abstract}
  We propose a viable design of a microwave two-photon threshold detector.
  In essence, the considered scheme is an extension of the existing single-photon detector---a Josephson photomultiplier (JPM)---an absorbing microwave detector based on a capacitively-shunted rf SQUID.
  To implement a two-photon threshold detector, we utilize a dimer of resonators---two lumped-element resonators interacting via an asymmetric dc SQUID, with one of the resonators capacitively coupled to the JPM.
  By specific tuning of the resonator frequencies and the external flux through the dc SQUID coupler, we engineer a non-perturbative two-photon coupling between the resonators. This coupling results in the coherent conversion of a photon pair from one resonator into a single photon in another resonator, enabling selective response to quantum states with at least two photons.
  We also consider an extended coupler design that allows on-demand \textit{in situ} switching of two-photon coupling.
  In addition, we propose the modified JPM design to improve its performance.
  Our calculations demonstrate that, for realistic circuit parameters, we can achieve more than $99\%$ fidelity of photon pair detection in less than 50 ns.
  The considered scheme may serve as a building block for the implementation of efficient photon-number-resolving detectors in circuit QED architecture.
\end{abstract}
\setcounter{page}{1}
\maketitle

\section{Introduction} \label{sec:intro}
Being an essential instrument in the quantum optics toolbox, the photodetector is extensively used in a variety of applications across the diverse domain of quantum technologies \cite{hadfield2009, eisaman2011, you2020}.
In particular, photodetectors are used in quantum-state engineering for the preparation of non-Gaussian states of light \cite{fiurasek2005, ourj2006, sperling2014, sonoyama2022, *sonoyama2023}.
They are also integral components required for the implementation of linear optics approaches to quantum computing \cite{knill2001, kok2007}, boson sampling \cite{broome2012, spring2013, tillmann2013}, and quantum-key distribution protocols \cite{shibata2014, koehler2018}.
However, these techniques are well-developed only for the infrared and visible light domains since there are several versatile realizations of highly efficient photodetectors \cite{hadfield2009, lita2022} operating in these spectral domains such as single-photon avalanche photodiodes \cite{yuan2008}, transition-edge sensors \cite{cabrera1998, lita2008}, and superconducting nanowire detectors \cite{natarajan2012, dauler2014, you2020, zadeh2021}.

When it comes to the detection of microwave quanta, the situation is drastically different.
The energy of microwave photons is 4-5 orders of magnitude lower than that of infrared and visible-light photons, making the detection of microwave photons a challenging problem.
The successful solution of this problem could allow one to harness the potential of the aforementioned quantum optics techniques in the microwave superconducting circuit QED architecture \cite{blais2004, gu2017, blais2021}, which has emerged as a highly promising hardware platform for building various devices for quantum information processing \cite{devoret2013, wendin2017, blais2020natphys}.

In addition, integrating on-chip microwave photodetectors helps address the challenge of interfacing between different temperature stages in circuit QED architecture---ranging from millikelvin to room temperature \cite{blais2021}.
Specifically, it could eliminate the need for bulky cascades of amplifiers and non-reciprocal elements ~\cite{jeffrey2014,walter2017,*heinsoo2018}, thereby improving the scalability of superconducting circuit QED systems.
This is essential for the practical implementation of error-corrected quantum computers.

A Josephson photomultipler (JPM) \cite{yfchen2011, oelsner2017, opremcak2018, opremcak2021} serves as an example of such a detector.
It features a simple on-chip design with a small physical footprint.
Technically, the JPM is a phase qubit~\cite{martinis2009qip} operating as a narrowband \emph{absorbing} detector.
Thus, it is fully compatible with the other superconducting circuit QED components.

Currently, two JPM designs are available.
The first one is a capacitively-shunted current-biased Josephson junction (CBJJ) \cite{yfchen2011,oelsner2017}.
The CBJJ has a washboard potential whose tilt is adjusted so that there are only a couple of levels in a local potential well, with the upper level close to the top of the barrier.
The probe photon with a frequency close to the transition frequency between these states drives the JPM to the upper level.
The excitation then rapidly tunnels through the barrier and ``slides'' down the washboard potential, resulting in a measurable voltage across the JJ, which is interpreted as a ``click''.
However, the switch to the voltage state produces an outburst of quasiparticles that hinders subsequent measurements, thereby reducing the measurement repetition rate of this JPM design and limiting its practical usability.

To address the limitations of the CBJJ-based scheme, another JPM design was devised~\cite{opremcak2018, opremcak2021}, which is based on a capacitively-shunted flux-biased rf SQUID~\cite{steff2006}.
By adjusting the external flux via the rf SQUID loop, the JPM potential is set to a highly asymmetric double-well configuration with a shallow well containing only a few levels and a deep well containing multiple levels.
Each potential well corresponds to a macroscopically distinct flux state.
An energy gap between the pair of the lowest levels in the shallow well is tuned to match the energy of the measured photons.
Unlike the CBJJ, in the flux-biased, the absorption of a photon does not completely switch the JPM to the voltage state but leads to a single interwell tunneling event accompanied by the flux-state change.
By determining the flux state of the JPM, one can infer whether the photon was absorbed or not.
In Refs.~\cite{opremcak2018, opremcak2021}, this type of JPM was used for high-fidelity dispersive readout of a transmon qubit.

The JPM operates as a single-photon detector (SPD), generating a macroscopic ``click'' response upon the absorption of a single photon per operating cycle.
More specifically, the JPM acts as a single-photon \emph{threshold} detector, trigerring the response when there is at least one photon in the measured electromagnetic mode.
In this paper, we propose the extension to the JPM that effectively transforms it into a two-photon threshold detector---one that, within a single operating cycle, absorbs a pair of photons and delivers a ``click" only when at least two photons are present in the measured mode.
In the dispersive qubit readout scheme with an absorbing photodetector~\cite{govia2014, sokolov2020pra},
a two-photon detector can provide higher readout accuracy~\cite{sokolov2016} compared to an SPD.
In addition, due to the pronounced selective response to quantum states containing at least two photons, this detector can be employed to keep a resonator mode state within the single-photon domain or to assess the fidelity of preparation of single-photon states by determining the multiphoton contribution.
Pairing a two-photon threshold detector with an additional SPD can provide limited photon-number resolution, enabling counting up to three photons in a single measurement iteration.
In a sequential counting scheme~\cite{stolyarov2023}, such an array detector could provide faster counting and enhanced performance due to fewer measurement iterations.

In Ref.~\cite{sokolov2020}, a feasible scheme of a two-photon detector based on a CBJJ-type JPM was proposed.
This scheme employs the anharmonicity of the JPM energy levels to engineer a \emph{perturbative} two-photon coupling between the JPM and the resonator mode.
However, this coupling is limited by the rather weak anharmonicity of the JPM potential.
For the JPM parameters used in Ref.~\cite{sokolov2020}, one requires several milliseconds to detect a pair of photons with high fidelity, which limits the practical use of such a detection scheme.

In this paper, we engineer a \emph{non-perturbative} two-photon interaction to implement fast high-fidelity two-photon detection.
For this purpose, we harness the inherent nonlinearity of a Josephson junction to implement an inductive-type nonlinear coupler~\cite{vrajitoarea2020, wang2024, ayyash2024, stolyarov2024arxiv} mediating the two-photon interaction.
We detail our design in the next section.

\begin{figure}[t!] 
    \centering
    \includegraphics{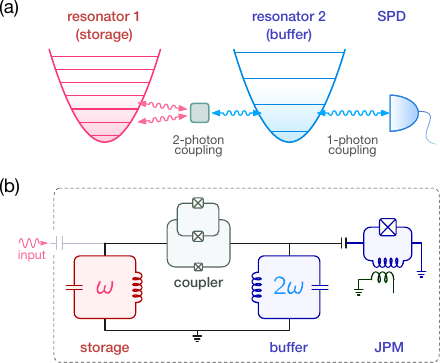}
    \caption{
    (a) Generic scheme for detection of photon pairs using a single SPD.
    The shaded square represents a nonlinear element that mediates two-photon coupling.
    (b) Schematics of the potential circuit QED setup implementing the considered approach to detection of microwave photon pairs.
    \label{fig:GeneralScheme}}
\end{figure}

\section{Principle of operation and Setup} \label{sec:Setup}

Let us first render the general idea underlying the considered design of a two-photon threshold detector without specifying the concrete details of its realization.
The scheme we consider is shown in Fig.~\ref{fig:GeneralScheme}(a).
To implement the two-photon detector, we propose using an SPD and two single-mode resonators referred to as the storage resonator and the buffer resonator.
The SPD is coupled to the buffer resonator through a resonant single-photon interaction, which converts a photon in the resonator mode into an excitation in the SPD, triggering a ``click''.

In addition to the SPD, the buffer resonator also interacts with the storage resonator---whose photons we aim to detect---through a \emph{nonlinear} inductive coupler.
The frequencies of the resonators are tailored so that the frequency of the buffer resonator is twice the frequency of the storage resonator.
This allows the nonlinear coupler to mediate the \emph{resonant} two-photon interaction, which leads to the coherent conversion of a photon pair in the storage resonator into a single photon in the buffer resonator and vice versa.
As a result, the considered model system exhibits substantially different behavior when two or more photons are present in the storage resonator than when it contains only one photon.
In the latter case, the buffer resonator remains in the vacuum state and the SPD does not click.\footnote{Here, we assume that the buffer resonator is initially in the vacuum state while the SPD is in its ground state. We also assume an idealized scenario without thermal photons in the system and false clicks of the SPD.}
On the contrary, when there are at least two photons in the storage resonator, two-photon coupling between the resonators leads to population of the buffer resonator and, thus, the SPD can absorb a photon and deliver a click.

Let us now outline the possible circuit QED implementation of the scheme described above.
Figure~\ref{fig:GeneralScheme}(b) illustrates the proposed realization.
The resonators are represented by the lumped-element $LC$ oscillators.
This type of microwave resonators offers a genuine single-mode behavior along with a small physical footprint.
The internal quality factors of state-of-the-art lumped-element resonators can exceed $10^6$~\cite{lilishi2022, crowley2023, dhundhwal2025arxiv}.
The resonators are bridged by a nonlinear coupler---either an asymmetric dc SQUID or its double-loop extension BiSQUID (see the details in Sec.~\ref{sec:BiSQUIDCoupler}).
The latter offers additional tunability through external magnetic fluxes.
Note that, in contrast to, \textit{e.g.}, Ref.~\cite{vrajitoarea2020}, which utilizes a flux-driven inductive coupler, the considered scheme employs \emph{static} fluxes similar to Refs.~\cite{wang2024, ayyash2024}.

The storage resonator can be coupled to control circuitry, allowing manipulation of its quantum state.
Alternatively, this resonator can be coupled to other quantum systems.
For example, it can be dispersively coupled to a qubit and driven by a classical tone to perform qubit readout.
Another option is to attach it to a waveguide for detecting itinerant photons \cite{besse2018, royer2018, lesc2020}.
However, in what follows, we leave these external elements beyond the scope, assuming that the quantum state to be measured is already prepared in the storage resonator.
The buffer resonator is capacitively coupled to a flux-biased JPM, which functions as an SPD.

In practical implementations, fabrication imperfections may compromise the  frequency matching condition required for the resonators.
Therefore, a certain degree of \textit{in situ} tunability of the resonators would be advantageous.
Here, we briefly outline some possible solutions that allow frequency adjustment.
Since the additional elements can potentially introduce an extra decoherence and reduce photon lifetime in the storage resonator, we assume it to be fixed-frequency lumped-element $LC$ oscillator.
Therefore, we would consider only the buffer resonator to be tunable.
One approach to frequency tuning involves shunting the fixed geometric inductance of the buffer resonator with either a dc SQUID~\cite{krantz2019, blais2021} or a gate-tunable JJ~\cite{larsen2015,*delange2015,strickland2023}.
These tunable nonlinear elements modify the full inductance of the resonator, enabling \textit{in situ} frequency adjustment.
As long as only moderate frequency variation is needed, the inductive energy of the geometric inductance can remain dominant.
This would ensure low sensitivity to fluctuations of the control flux or voltage.
Additionally, in the dc SQUID variant, introducing asymmetry in the SQUID helps reduce sensitivity to flux noise \cite{koch2007, krantz2019}.
Another approach is to replace the fixed geometric inductance with the tunable inductance realized by an array of dc SQUIDs~\cite{wang2019prx, *dassonneville2020prx}.
Alternatively, the buffer resonator can be substituted with a flux-tunable split transmon~\cite{koch2007, barends2013} or a voltage-controlled gatemon~\cite{larsen2015,*delange2015, casparis2018}.
Keeping these modifications in mind, we omit them in this paper for the sake of conciseness and focus on the fixed-frequency design shown in Fig.~\ref{fig:GeneralScheme}(b).

In the considered setup, the rf-SQUID JPM serves as an SPD.
However, it is important to emphasize that the considered scheme for detecting pairs of microwave photons is also compatible with the CBJJ-type JPMs, as well as with other types of microwave SPDs that support capacitive coupling to the lumped-element resonator, such as transmon-based detectors~\cite{besse2018, dass2020prappl, lesc2020, balembois2024}.
This versatility is one of the key advantages of the presented approach.

In the next two sections, we discuss the key elements of the proposed detection scheme.
In Sec.~\ref{sec:TwoPhotonCoupling}, we provide a detailed analysis of the SQUID-coupled resonators subsystem.
We outline the design and operation principle of the rf-SQUID JPM in Sec.~\ref{sec:JPM}.

The remainder of the paper is organized as follows.
In Sec.~\ref{sec:MasterEquation}, we discuss the master equation describing the evolution of the quantum state of the system under consideration.
The performance of the considered scheme is analyzed in Sec.~\ref{sec:Performance}.
We summarize our results in Sec.~\ref{sec:summ}.
Detailed derivations are delegated to Appendixes.

\section{Two-photon coupling} \label{sec:TwoPhotonCoupling}
In this section, we elaborate on the implementation of a tunable two-photon coupling between the resonators using a nonlinear inductive coupler.

\subsection{Asymmetric SQUID coupler} \label{sec:AsymmetricSQUID}
Consider two single-mode resonators that are inductively coupled via an asymmetric dc SQUID threaded by the static external flux $\Phi_\cpl$. 
The JJs comprising the coupler are characterized by critical currents $I^\cpl_0$ and $\alpha I^\cpl_0$, where $\alpha \geq 0$, and the corresponding self-capacitances $C^\cpl_\JJ$ and $C^\cpl_{\alpha \JJ}$.
This ``dimer'' of resonators is described by the Hamiltonian given by
\begin{equation} \label{eq:HamiltonianResRes}
    \opH_{\res\res} = \sum_{j=1}^2 \opH_{\res j} + \opH^\mathrm{int}_{\res\res}.
\end{equation}
Here, $\opH_{\res j}$ stands for the Hamiltonian of the $j$th resonator expressed as
\begin{equation} \label{eq:ResonatorHamiltonian}
    \opH_{\res j} = \hbar\omega_{\osc j} \hat a^\dag_j \hat a_j - \hbar \sum_{k\geq 3} (-1)^{(j - 1) k} \mathcal{G}_{k\,j} (\hat a^\dag_j + \hat a_j)^k,
\end{equation}
where operator $\hat a_j$ annihilates a photon in the mode of the $j$th resonator.
Hereinafter, the indices $j=1,2$ correspond to the storage and buffer resonator, respectively.
The above Hamiltonian describes a quantum nonlinear oscillator of frequency
\begin{equation} \label{eq:ResonatorFrequency}
    \omega_{\osc j} = \frac{1}{\hbar}\sqrt{8 \tilde{E}_{Cj} \tilde{E}_{Lj}},
\end{equation}
where $\tilde{E}_{Cj}$ and $\tilde{E}_{Lj}$ are the capacitive and inductive energies of the $j$th resonator \emph{renormalized} due to the SQUID coupler.
The detailed derivation of the Hamiltonian~\eqref{eq:HamiltonianResRes} as well as the definitions of the renormalized energies are given in Appendix~\ref{sec:HamiltonianDerivation}.

The resonators acquire their nonlinearity from the coupling SQUID, which is an inherently nonlinear element.
The strength of the $k$th order nonlinearity is
\begin{equation} \label{eq:ResonatorNonlinearity}
 \mathcal{G}_{k\,j} = \frac{E^\mathrm{c}_{\JJ\,\mathrm{eff}}}{\hbar} \frac{\varphi_{\zpf j}^k u_k}{k!},
\end{equation}
where $\varphi_{\zpf j} = (2\tilde{E}_{Cj}/\tilde{E}_{Lj})^{1/4}$ is the characteristic magnitude of the zero-point fluctuations of the condensate phase drop over the $j$th resonator.
The \emph{effective} flux-dependent Josephson energy of the coupler $E^\mathrm{c}_{\JJ\,\mathrm{eff}}$ is determined as~\cite{koch2007, sun2023}
\begin{equation} \label{eq:EJ_eff}
   E^\mathrm{c}_{\JJ\,\mathrm{eff}} = E^\cpl_\JJ \, (1 + \alpha) \sqrt{\cos^2 \left(\pi\frac{\Phi_\cpl}{\Phi_0}\right) + \xi^2 \sin^2\left(\pi\frac{\Phi_\cpl}{\Phi_0}\right)},
\end{equation}
where $E^\cpl_\JJ = I^\cpl_0 \Phi_0/2\pi$ is the Josephson energy of the junction in the coupler, $\xi = (1-\alpha)/(1+\alpha)$, and $\Phi_0 = h/2e$ denotes the magnetic flux quantum.

The coefficient $u_k$ is determined as
\begin{equation} \label{eq:uk}
 u_k = (-1)^{\lceil k/2 \rceil}
 \left\{
 \begin{array}{ll}
       \cos\delta, \quad \textrm{for even $m$}, \\
       \sin\delta, \quad \textrm{for odd $m$},
 \end{array}
 \right.
\end{equation}
where the ceiling function $\lceil k/2 \rceil$ denotes the lowest integer greater than or equal to $k/2$.
The equilibrium phase difference over the coupler $\delta$ is given by
\begin{equation} \label{eq:DeltaDefinition}
    \delta = \phi^\mathrm{min}_1 - \phi^\mathrm{min}_2 - \vartheta.
\end{equation}
Here, $\phi^\mathrm{min}_1$ and $\phi^\mathrm{min}_2$ are the equilibrium phases of the two-dimensional potential energy of the circuit (see Appendix~\ref{sec:HamiltonianDerivation}), while the phase shift $\vartheta$ is determined as
\begin{equation} \label{eq:PhaseShift}
  \vartheta = \pi\frac{\Phi_\cpl}{\Phi_0} + \arctan\left[\xi \tan\left(\pi\frac{\Phi_\cpl}{\Phi_0}\right)\right].
\end{equation}

The interaction between the resonators is covered by
\begin{equation} \label{eq:InteractionHamResRes}
   \begin{split}
     \opH^\mathrm{int}_{\res\res} = & \, - \hbar g_c (\hat a^\dag_1 - \hat a_1)(\hat a^\dag_2 - \hat a_2) \\
     & \, + \hbar \sum_{k \geq 2} \sum_{l=1}^{k-1} g_{k-l \, l} (\hat a^\dag_1 + \hat a_1)^{k-l} (\hat a^\dag_2 + \hat a_2)^{l}.
   \end{split}
\end{equation}
The first term describes the static capacitive coupling of strength
\begin{equation} \label{eq:CapacitiveCoupling}
 g_c = \frac{E^\cpl_C}{\hbar} n_{\zpf 1} n_{\zpf 2},
\end{equation}
where $n_{\zpf j} = (\tilde{E}_{Lj}/32\tilde{E}_{Cj})^{1/4}$ stands for the magnitude of the zero-point fluctuations of the charge on $j$th resonator capacitance, and $E^\cpl_C$ is a capacitive coupling energy (see Appendix \ref{sec:HamiltonianDerivation}).
The sum in the bottom line of Eq.~\eqref{eq:InteractionHamResRes} represents all the interactions stemming from the nonlinear inductance of the coupler.
Coupling strength $g_{k-l \, l}$ is determined as
\begin{equation} \label{eq:gklCoupling}
  g_{k-l \, l} = (-1)^{l+1} \frac{E^\mathrm{c}_{\JJ\,\mathrm{eff}}}{\hbar} \frac{\varphi_{\zpf 1}^{k-l} \varphi_{\zpf 2}^l u_k}{(k-l)! \, l!}.
\end{equation}
In the further analysis, we assume that the couplings between the resonators are much weaker than their frequencies, $g_c, |g_{k-l \, l}| \ll \omega_{\osc 1}, \omega_{\osc 2}$.

In the interaction Hamiltonian, the term $\propto (\hat a^\dag_1 + \hat a_1)(\hat a^\dag_2 + \hat a_2)$ embodies the linear inductive coupling with strength $g_{11} = \varphi_{\zpf 1}\varphi_{\zpf 2} u_2/\hbar$, which enables the single-photon exchange between the resonators.
The higher-order terms ($k \geq 3$) in the Hamiltonian~\eqref{eq:InteractionHamResRes} describe various non-dipolar interactions that emerge from coupler nonlinearity.
In particular, terms $\propto (a^{\dag \, k-l}_1 a^l_2 + \mathhc)$ describe the degenerate $k$-wave mixing corresponding to the conversion of $k-l$ photons in the storage resonator to $l$ photons in the buffer resonator and vice versa.
We can bolster up one of these processes to make it dominant.
By specifically tuning the frequencies of the resonators, we bring one of the couplings resonant while keeping the others strongly off-resonant.
The system’s dynamics are governed primarily by the resonant interaction, with the off-resonant processes entering as perturbative corrections.
As outlined in Sec.~\ref{sec:Setup}, the proposed two-photon detector scheme requires a coupling that converts two photons in the storage resonator to one photon in the buffer resonator being the dominant interaction between the resonators.
To implement such a coupling regime, we adjust the frequencies of the resonators to satisfy the two-photon resonance condition $\omega_{\osc 1} \approx \omega_{\osc 2}/2$.

Equations~\eqref{eq:uk} and~\eqref{eq:gklCoupling} suggest that by adjusting the coupler bias to satisfy the condition
\begin{equation} \label{eq:CrossKerrFree}
	\delta = \pm \left(\mathsf{n} + \frac{1}{2}\right)\pi, \quad \mathsf{n} \in \mathbb{Z},
\end{equation}
one switches off all redundant even-order inductive couplings in the Hamiltonian~\eqref{eq:InteractionHamResRes}.
In particular, the cross-Kerr--type interactions $\propto \hat a^{\dag (k-l)}_1 \hat a^{(k-l)}_1 \hat a^{\dag l}_2 \hat a^l_2$, where $k\geq 2$ and $1\leq l <k$, vanish in this regime.
These interactions induce photon-number--dependent shifts in the resonator frequencies.
Since any information about the number of photons in the measured state is \emph{a priori} unavailable, these couplings can deteriorate the detection efficiency, especially, for states with a large number of photons.

In what follows, we assume the system is set in the regime with vanishing even-order couplings determined by the condition~\eqref{eq:CrossKerrFree}.
Then, we omit the off-resonant couplings in the interaction Hamiltonian~\eqref{eq:InteractionHamResRes}.
Treating these terms as perturbations would lead to onset of cross-Kerr--type couplings and shifts in the resonator frequencies.
However, for the considered circuit parameters, we estimate that these renormalizations are much weaker than the dominant interactions, namely, two-photon coupling between the resonators and single-photon coupling between the buffer resonator and the JPM.
Among the remaining odd-order resonant terms, we keep the lowest-order one assuming that $\varphi_{\zpf j}^2 \ll 1$.
This yields the interaction Hamiltonian
\begin{equation} \label{eq:HamIntApprox}
  \begin{split}
    \opH^\mathrm{int}_{\res\res} \approx {}& \hbar g_{21} (\hat a^{\dag 2}_1 \hat a_2 + \hat a^\dag_2 \hat a^2_1),
  \end{split}
\end{equation}
which represents the dominant two-photon coupling with strength
\begin{equation}
	g_{21} = \frac{1}{2}\frac{E^\cpl_{\JJ\,\eff}}{\hbar} \varphi^2_{\zpf 1} \varphi_{\zpf 2} \sin\delta.
\end{equation}
Note that the analogous expression was derived in Ref.~\cite{stolyarov2024arxiv} for the two-photon coupling strength between a lumped-element $LC$ oscillator and an rf SQUID.

\begin{figure}[t!] 
	\centering
	\includegraphics{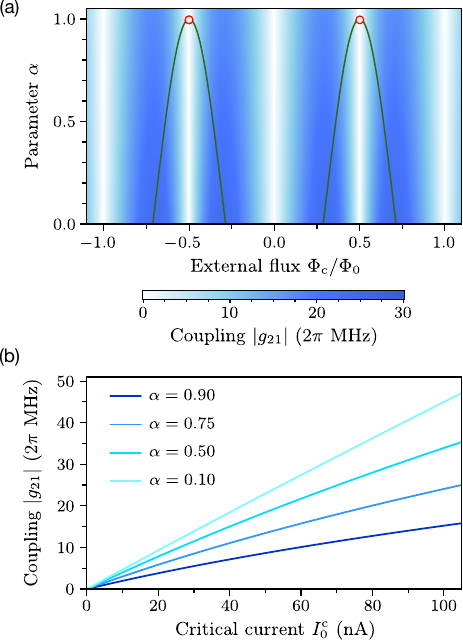}
	\caption{(a) Dependence of the two-photon coupling strength $|g_{21}|$ on the external flux $\Phi_\cpl$ via the coupler and the asymmetry parameter $\alpha$.
		Solid lines correspond to combinations of $\textstyle\Phi_\cpl$ and $\alpha$ providing the the odd-parity coupling regime.
		Circles correspond to vanishing Josephson coupling $E^\cpl_{\JJ\,\mathrm{eff}} = 0$.
		Here we set $I^\cpl_0 = 50\,\mathrm{n A}$ ($E^\cpl_{\JJ}/h = 24.8\,\mathrm{GHz}$).
		(b) Dependence of the two-photon coupling strength $|g_{21}|$ in the the odd-parity coupling regime on the critical current $I^\cpl_0$ for different values of $\alpha$.
		Parameters of the resonators used for calculations are as follows: $C_1 = 1\,\mathrm{pF}$ ($E_{C1}/h = 19.4\,\mathrm{MHz}$), $L_1 = 1\,\mathrm{nH}$ ($E_{L1}/h = 163.2\,\mathrm{GHz}$), $C_2 = 0.5\,\mathrm{pF}$ ($E_{C2}/h = 38.7\,\mathrm{MHz}$), and $L_2 = 0.5\,\mathrm{nH}$ ($E_{L2}/h = 326.4\,\mathrm{GHz}$).
	\label{fig:ASQUIDResults}}
\end{figure}

\begin{figure*}[t!] 
	\centering
	\includegraphics{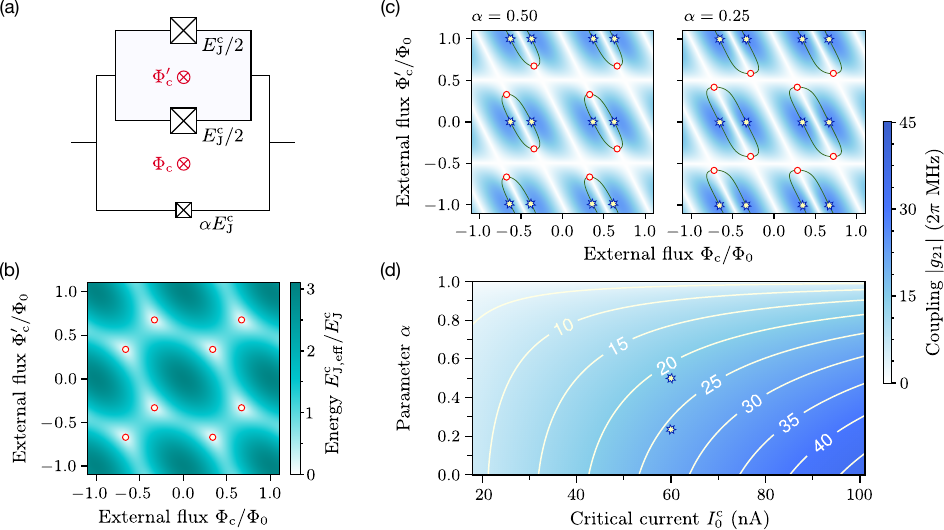}
	\caption{(a) Circuit diagram of the BiSQUID coupler.
		(b) Effective Josephson energy $E^\cpl_{\JJ\,\mathrm{eff}}$ of the BiSQUID as a function of the external fluxes $\Phi_\cpl$ and $\Phi^\prime_\cpl$ threading through its loops for $\alpha = 0.5$.
		Empty hexagons mark $E^\cpl_{\JJ,\mathrm{eff}} = 0$.
		(c) Dependence of the two-photon coupling strength $|g_{21}|$ on the external fluxes $\Phi_\cpl$ and $\Phi^\prime_\cpl$ via the BiSQUID loops for the different values of $\alpha$ indicated above the corresponding plot.
		Solid lines indicate the values of the external fluxes providing the the odd-parity  coupling.
		Stars mark the external fluxes providing the maximum two-photon coupling in the the odd-parity coupling regime.
		Circles correspond to the fluxes for which the Josephson coupling vanishes $E^\cpl_{\JJ\,\mathrm{eff}} = 0$.
		Here, we set $I^\cpl_0 = 60\,\mathrm{nA}$ ($E^\cpl_\JJ/h = 29.8\,\mathrm{GHz}$).
		(d) Dependence of the maximum two-photon coupling strength on the parameter $\alpha$ and the critical current $I_0^\cpl$.
		Stars correspond to the coupler parameters used in Fig.~\ref{fig:BiSQUIDResults}(c).
		Parameters of the resonators are the same as used in Fig.~\ref{fig:ASQUIDResults}.
		\label{fig:BiSQUIDResults}}
\end{figure*}

To proceed, similar to the interaction Hamiltonian, we keep only the lowest-order nonlinear term in the resonator Hamiltonian~\eqref{eq:ResonatorHamiltonian}, which leads to
\begin{equation} \label{eq:ResonatorHamiltonianApprox}
    \opH_{\res j} \approx \hbar\omega_{\osc j} \hat a^\dag_j \hat a_j + (-1)^{j} \hbar\mathcal{G}_{3 j}(\hat a^\dag_j + \hat a_j)^3.
\end{equation}
Next, we diagonalize this Hamiltonian via the Schrieffer-Wolff transformation \cite{schrieffer1966, bravyi2011, zhu2013}: $\opH_{\res j} \rightarrow \hat{\cH}_{\res j} = \hat{U}^\dag_\res \opH_{\res j} \hat{U}_\res$.
The unitary operator $\hat{U}_\res$ is given by~\cite{hillmann2022, stolyarov2023}:
\begin{equation} \label{eq:UnitaryTransformLambda}
 \hat{U}_\res = \exp\left[\sum_{j=1}^2 \varLambda_j\left(\frac{\hat a^{\dag 3}_j}{3} + 3 \hat a^{\dag 2}_j \hat a_j + 3 \hat a^\dag_j - \mathhc\right)\right],
\end{equation}
where $\varLambda_j = \mathcal{G}_{3 j}/\omega_j$, with $\varLambda_j \ll 1$ for considered  system parameters.
Using the Baker-Campbell-Hausdorff relation $\ee^{-\hat A} \hat B \ee^{\hat A} = \hat B + [\hat B, \hat A] + \frac{1}{2!} [[\hat B, \hat A],\hat A] + \ldots$ and keeping the terms up to the first order in $\varLambda_j$, one obtains
\begin{equation} \label{eq:HamKerrOsc}
  \hat{\cH}_{\res j} \approx \hbar \omega_j \hat a^\dag_j \hat a_j - \hbar K_j \hat a^{\dag 2}_j \hat a^2_j.
\end{equation}
This expression corresponds to the Hamiltonian of the self-Kerr nonlinear oscillator, where $\omega_j = \omega_{\osc j} - 2 K_j$ is its renormalized frequency, while $K_j$ stands for the self-Kerr nonlinearity strength determined as $K_j = 30\cG^2_{3j}/\omega_{\osc j}$.

Dropping the irrelevant off-resonant terms, the transformation~\eqref{eq:UnitaryTransformLambda} leaves the  interaction Hamiltonian~\eqref{eq:HamIntApprox} unchanged, $\hat U_\res^\dag \hat H^\mathrm{int}_{\res\res} \hat U_\res \approx \hat H^\mathrm{int}_{\res\res}$.
Putting all terms together, the effective Hamiltonian describing the coupled resonators in the odd-parity interaction regime reads as
\begin{equation} \label{eq:hamrr_qnt4}
	\begin{split}
		\hat{\cH}_{\res\res} \approx \hbar \sum_{j=1}^2 \left(\omega_j \hat a^\dag_j \hat a_j - K_j \hat a^{\dag 2}_j \hat a^2_j\right) + \hbar g_{21} (\hat a^{\dag 2}_1 a_2 + \hat a^\dag_2 \hat a^2_1).
	\end{split}
\end{equation}
Note that for our circuit parameters, the self-Kerr nonlinearities of the resonators are significantly weaker than the two-photon coupling, \textit{i.e.}, $K_j \ll |g_{21}|$.

Figure~\ref{fig:ASQUIDResults}(a) demonstrates the dependence of the two-photon coupling strength $|g_{21}|$ on the coupler parameters: SQUID asymmetry $\alpha$ and the external flux $\Phi_\cpl$.
The solid lines indicate the combination of parameters $\alpha$ and $\Phi_\cpl$, for which the odd-parity  coupling regime is achieved.
The results of calculations shown in Fig.~\ref{fig:ASQUIDResults}(b) suggest that stronger asymmetry (lower $\alpha$) of the coupling SQUID yields stronger two-photon coupling.
In the limiting case of a symmetric SQUID ($\alpha = 1$), the even-order couplings vanish for $\Phi_\cpl = (\mathsf{n} + \tfrac{1}{2})\Phi_0$.
However, the requisite two-photon coupling also vanishes for this flux bias since the Josephson coupling zeroes $E^\cpl_{\JJ\,\eff} = 0$.

\subsection{BiSQUID coupler} \label{sec:BiSQUIDCoupler}
Above, we have shown that the two-photon coupling required for implementing the two-photon detector considered in Sec.~\ref{sec:Setup} can be achieved by bridging the resonators via the asymmetric dc SQUID.
However, this design generally does not have a knob to switch on/off the inductive coupling between the resonators \emph{in situ} on demand, although this feature may be beneficial for some applications.
The only way to completely switch off the dc SQUID coupler (\textit{i.e.}, ensure that both even- and odd-order couplings vanish) is to make it symmetric, $\alpha=1$, as shown in Fig.~\ref{fig:ASQUIDResults}(a).
However, this configuration would restrict the desired functionality of the coupler.
Specifically, it would prevent achieving the odd-parity coupling regime during detector operation.
To overcome this limitation, we introduce a simple modification of the asymmetric SQUID coupler that addresses this issue.
Namely, the replacement of the larger JJ (with the critical current $I^\cpl_0$) with a symmetric SQUID (composed of two JJs with the critical current $I^\cpl_0/2$) threaded by the external flux $\Phi^\prime_\cpl$.
The obtained BiSQUID \cite{griesmar_phdthesis, *peyruchat_phdthesis, peyruchat2024} operates here essentially as an asymmetric SQUID with tunable asymmetry.
Circuit diagram of the considered BiSQUID coupler is shown in Fig.~\ref{fig:BiSQUIDResults}(a).

Indeed, the introduced symmetric SQUID acts here as a single JJ with a flux-tunable Josephson energy $E^\cpl_\JJ \cos(\pi \Phi^\prime_\cpl/\Phi_0)$ \cite{tinkham2004, krantz2019}.
Thus, the coupler asymmetry is now controlled by the flux $\Phi^\prime_\cpl$.
By adjusting this flux to satisfy $\Phi'_\cpl/\Phi_0 = \arccos(\alpha)/\pi$, one sets the BiSQUID in the regime of a symmetric SQUID whose Josephson energy can be zeroed by setting ${\Phi_\cpl = \pm (\mathsf{n} + \frac{1}{2}) \Phi_0 - \frac{1}{2}\Phi^\prime_\cpl}$.
This implies that one can \emph{on-demand} switch off the inductive coupling mediated by the BiSQUID, leaving only the coupling attributed to the self-capacitances of the JJs.
However, the capacitive coupling is strongly off-resonant, so the resonators are essentially decoupled in this setting.

\begin{figure*}[t!] %
	\centering
	\includegraphics{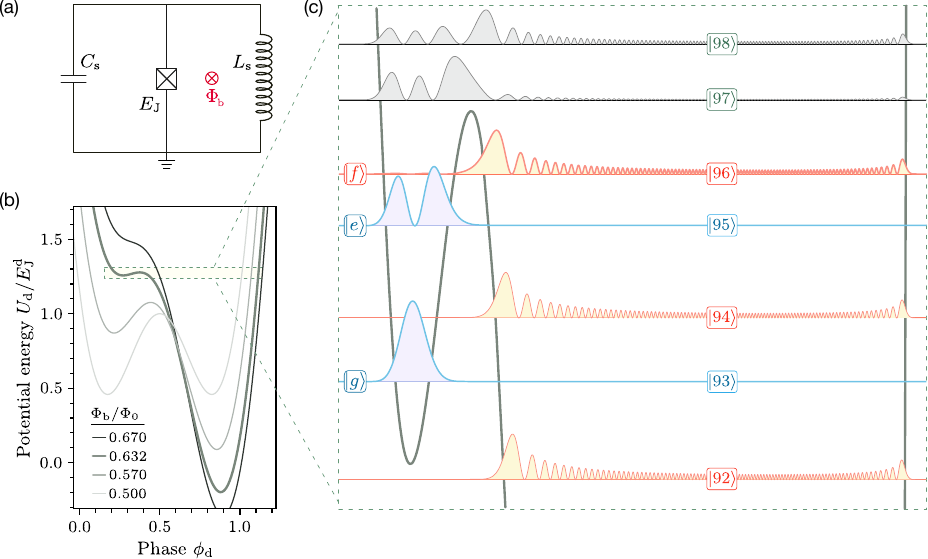}
	\caption{(a) Lumped-element circuit diagram of the JPM represented by the flux-biased capacitively-shunted rf-SQUID.
	(b) Configurations of the JPM potential for the different values of the bias flux $\Phi_\mathrm{b}$.
		(c) Zoomed view of the region of the JPM potential [indicated by the dashed square in Fig.~\ref{fig:JPMSchemePotential}(b)] for $\Phi_\mathrm{b} = 0.6316\Phi_0$ in the vicinity of the shallow well for the highly asymmetric configuration with only two eigenlevels localized in the left (shallow) well, while there are 94 eigenlevels localized in the right (deep) well.
        We demonstrate the JPM eigenlevels and the square moduli of the corresponding wavefunctions.
        Note that in what follows, c.f., Sec.~\ref{sec:JPMModel}, we relabel the pair of eigenstates $|93\rangle$ and $|95\rangle$ localized in the shallow well as $|g\rangle$ and $|e\rangle$, and the eigenlevel $|96\rangle$ localized in the deep well as $|f\rangle$.
        The JPM parameters used for calculations are shown in Table~\ref{tab:JPMParams}.
	\label{fig:JPMSchemePotential}}
\end{figure*}

The effective Josephson energy of the considered BiSQUID for the arbitrary values of fluxes $\Phi_\cpl$ and $\Phi^\prime_\cpl$ is determined as
\begin{equation} \label{eq:EJBiSQUID}
 \begin{split}
   E^\cpl_{\JJ\,\mathrm{eff}} = & \, E^\cpl_\JJ \left[\cos\left(\pi\frac{\Phi'_\cpl}{\Phi_0}\right) + \alpha\right] \\
    & \, \times \sqrt{\cos^2\left(\pi \frac{\Phi_\cpl + \frac{1}{2}\Phi^\prime_\cpl}{\Phi_0}\right) + \xi^2 \sin^2\left(\pi \frac{\Phi_\cpl + \frac{1}{2}\Phi^\prime_\cpl}{\Phi_0}\right)},
 \end{split}
\end{equation}
where $\xi$ is now defined as
\begin{equation} \label{eq:BiSQUIDxi}
 \xi = \frac{\cos\left(\pi \frac{\Phi'_\cpl}{\Phi_0}\right) - \alpha}{\cos\left(\pi \frac{\Phi'_\cpl}{\Phi_0}\right) + \alpha}.
\end{equation}
In the case of the BiSQUID coupler, the phase shift $\vartheta$ in Eq.~\eqref{eq:ClassicalHamResRes} acquires the form
\begin{equation} \label{eq:PhaseShiftBiSQUID}
    \vartheta = \pi \frac{\Phi_\cpl + \frac{1}{2}\Phi^\prime_\cpl}{\Phi_0} + \arctan\left[\xi\tan\left(\pi \frac{\Phi_\cpl + \frac{1}{2}\Phi^\prime_\cpl}{\Phi_0}\right)\right],
\end{equation}
where $\xi$ is given by Eq.~\eqref{eq:BiSQUIDxi}.
For $\Phi^\prime_\cpl = 0$, Eqs.~\eqref{eq:EJBiSQUID} and \eqref{eq:PhaseShiftBiSQUID} reduce to Eqs.~\eqref{eq:EJ_eff} and \eqref{eq:PhaseShift}, respectively, which describe the asymmetric SQUID.
Dependence of the effective Josephson energy of the BiSQUID given by Eq.~\eqref{eq:EJBiSQUID} on the external fluxes threading via its loops is shown in Fig.~\ref{fig:BiSQUIDResults}(b).

Using the expressions for the effective Josephson energy $E^\cpl_{\JJ,\mathrm{eff}}$ and the phase shift $\vartheta$ given by Eqs.~\eqref{eq:EJBiSQUID} and \eqref{eq:PhaseShiftBiSQUID}, respectively, in Eqs.~\eqref{eq:gklCoupling}, \eqref{eq:uk}, and \eqref{eq:kirch}, we determine the couplings between the resonators bridged by the BiSQUID.
Figure~\ref{fig:BiSQUIDResults}(c) shows the dependence of the two-photon coupling strength $|g_{21}|$ on the external fluxes $\Phi_\cpl$ and $\Phi^\prime_\cpl$ threading via the BiSQUID loops.
The plot demonstrates that the BiSQUID coupler allows the implementation of the odd-parity  coupling regime as well as the regime with \emph{all} inductive couplings switched off.
For given values of $\alpha$ and $I^\cpl_0$, the maximum two-photon coupling strength is achieved when $\Phi^\prime_\cpl = 0$.
The dependence of the maximum $|g_{21}|$ on the critical current $I^\cpl_0$ and the parameter $\alpha$ is shown in Fig.~\ref{fig:BiSQUIDResults}(d).

\section{Josephson photomultiplier} \label{sec:JPM}

In this section, we discuss the design and the principle of operation of the JPM, which serves as the SPD in our scheme of detection of photon pairs.
Here, we also provide the model describing the photomultiplier.

\subsection{Generic scheme and operation principle} \label{sec:JPM_oper}
The JPM design under consideration consists of an rf SQUID loop shunted by a capacitance $C_\mathrm{S}$.
The rf SQUID is composed of an inductance $L_\mathrm{S}$ and a JJ characterized by the critical current $I^\pd_0$, which corresponds to the Josephson energy $E_\JJ = I^\pd_0 \Phi_0/2\pi$, and the self-inductance $C_\JJ$.
The inductance comprising the rf SQUID can be either a geometric inductance, as in Refs.~\cite{opremcak2018, opremcak2021}, a high-kinetic inductance superconducting nanowire \cite{peltonen2018, niepce2019}, or a chain of multiple JJs, similar to those routinely used in fluxonium-type artificial atoms~\cite{manucharyan2009, nguyen2019, bao2022}.
The rf SQUID loop of the JPM is threaded by the magnetic flux $\Phi_\mathrm{b}$ generated by the external flux bias line (FBL).

The replacement of the JJ with a dc SQUID controlled via a dedicated FBL \cite{sun2023} or with a gate-tunable JJ \cite{larsen2015,*delange2015,strickland2023} provides an additional means to manipulate the JPM frequency~\cite{shnyrkov2018}.
In addition, the inductance can be made tunable by using either an array of multiple dc SQUIDs~\cite{krupko2018} controlled by a common FBL~\cite{puertasmartinez2019} or a gate-tunable kinetic inductance~\cite{splitthoff2022,*splitthoff2024}.
However, for the sake of simplicity, we focus here on the scheme with a single JJ and a static inductance, similar to that experimentally demonstrated in Refs.~\cite{opremcak2018, opremcak2021}.
Figure~\ref{fig:JPMSchemePotential}(a) shows the circuit diagram of the considered JPM.

\begin{table}[t!]
	\caption{Circuit parameters of the JPM and the corresponding energies (given in frequency units) used for calculations. \label{tab:JPMParams}}
	\begin{centering}
		\begin{ruledtabular}
            \newcolumntype{2}{D{.}{.}{2}} 
                \renewcommand{\arraystretch}{1.1}
			\begin{tabular}{lcc}
                    \multicolumn{2}{l}{\textbf{Circuit parameters:}} \\
				    \, JJ critical current & $I_0$ & 2.5 $\mu$A \\
                    \, Inductance & $L_S$ & 300 pH\\
                    \, Total capacitance\footnote{Taking the JJ capacitance density of 50 fF/$\mu \mathrm{m^2}$ and its critical current density of 1 $\mu$A/$\mu \mathrm{m^2}$, we estimate the self-capacitance of the JPM junction to be of $C^\pd_\JJ \approx 125 \, \textrm{fF}$.} & $C_\pd$ & 405 fF \\
                    \hline
                    \multicolumn{2}{l}{\textbf{Energies (in freq. units):}} \\
                    \, Josephson energy  & $E^\pd_\JJ/h$ & 1243.4 GHz \\
                    \, Inductive energy  & $E^\pd_L/h$ & 544.0 GHz \\
                    \, Capacitive energy & $E^\pd_C/h$ & 47.9 MHz
			\end{tabular}
		\end{ruledtabular}
	\end{centering}
\end{table}

In the rf-SQUID--based photomultiplier, the shunting inductance modulates the cosine-shaped potential of the JJ with the quadratic term.
The external flux through the rf SQUID loop determines the overall shape of the JPM potential as shown in Fig.~\ref{fig:JPMSchemePotential}(b).
With proper adjustment of the inductance $L_\mathrm{S}$ and the JJ critical current $I_0$, the JPM potential can exhibit either one or two wells, depending on the flux $\Phi_\mathrm{b}$.
To probe the coupled resonator photons [see Fig.~\ref{fig:GeneralScheme}(b)], the flux $\Phi_\mathrm{b}$ is adjusted such that the JPM potential has an asymmetric double-well configuration with just a few eigenstates localized in a shallow well, and multiple eigenstates localized in the deep well.
The energy gap between the two lowest eigenstates localized in the shallow well is tuned to be close to the energy of the resonator photon.
In this case, the photon induces the transition between these states.
If the upper eigenstate in the left well lies close to the top of the potential barrier separating the wells, the excitation can relax to the eigenstates in the deep well and then rapidly ``fall'' down to the bottom eigenstate dissipating its energy~\cite{opremcak2018, shnyrkov2023}.
The described process entails a classically distinguishable change of the flux state of the JPM, which is interpreted as a ``click'' event.
Thus, by reading the flux state of the JPM, one can infer whether the JPM has ``clicked'', \textit{i.e.}, absorbed a photon, or not.

There are several techniques for the readout of the JPM flux state.
Although we do not address any particular implementation of the JPM readout in our consideration, we briefly outline these techniques for consistency.
Transition from the left well to the right well results in the change of the direction of current circulating by the JPM loop which yields the change of the magnetic flux field generated by the JPM.
This change of the JPM flux state can be detected by an rf SQUID, which is weakly coupled to the JPM~\cite{shnyrkov2023}.
Additionally, the JPM flux state can be probed by ballistic fluxons propagating through an underdamped Josephson transmission line interfaced with single-flux quantum logic~\cite{howington2019, howington2019thesis}.
The flux generated by the JPM induces a weak current in the Josephson transmission line affecting the propagating fluxon, which acquires a conditional time delay depending on the JPM flux state \cite{averin2006, fedorov2007}.
By determining the delay of the fluxon using the single-flux quantum logic, one infers the JPM state. 
Another technique, demonstrated in Refs.~\cite{opremcak2018, opremcak2021}, relies on the fact that in the double-well configuration, each potential well is characterized by a different plasma frequency.
By probing the JPM with a weak microwave drive at a frequency corresponding to the plasma frequency of the deep well, one can determine whether the change of the JPM flux state occurred with high ($>99.9\%$) fidelity.

Noteworthy, even if the JPM has not absorbed a photon, there is still a nonzero probability that it will deliver a ``click'' caused by the relaxation from the bottom eigenstate in the shallow well to the eigenstates in the deep well.
Such false counts deteriorate the performance of the photodetector.
In particular, they reduce the resolution of the photocounting schemes based on the repeated measurement of the resonator mode by a single detector~\cite{stolyarov2023}.

\subsection{Design modifications} \label{sec:JPMModifications}
To improve the performance of the JPM, in particular, to reduce the false count rate, we propose some modifications of its design compared to that demonstrated in Refs.~\cite{opremcak2018, opremcak2021}.
The superfluous tunneling from the bottom eigenstate in the left well can be suppressed by making the left well deeper, thus making a potential barrier wider.
However, this also suppresses the requisite tunneling from the first excited state, which results in a longer detection time, \textit{i.e.}, slower operation.
To overcome this trade-off between speed of operation and false count rate, we exploit an eigenstate localized in the deep well.
We configure the JPM potential so that the left well accommodates a pair of energy levels deep within the well, reducing the false count rate.
At the same time, we apply a coherent drive tone to the transition between the upper energy level in the left well and the upper level in the right well.
This ensures that when the JPM absorbs a photon from the resonator mode, the drive induces a transition to the upper eigenstate in the right (deep) well, and the excitation then rapidly ``rolls'' down to the bottom of the deep well.
This approach is similar to that used in Ref.~\cite{simmonds2004prl} for destructive readout of a phase qubit.
The discussed configuration of the JPM eigenstates in the vicinity of the potential barrier is shown in Fig.~\ref{fig:JPMSchemePotential}(c).

\begin{figure}[t!] %
	\centering
	\includegraphics{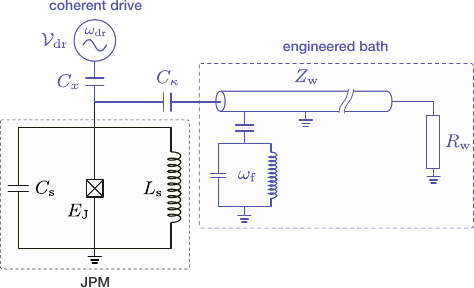}
	\caption{Modified JPM design. The JPM couples to the engineered bath represented by a waveguide of impedance $Z_\mathrm{w}$ terminated by the resistive element $R_\mathrm{w}$. The waveguide also couples to the filter resonator of frequency $\omega_\mathrm{f}$.
    The JPM is driven by the coherent tone of frequency $\omega_\mathrm{dr}$.}
    \label{fig:ModifiedJPM}
\end{figure}

Following Ref.~\cite{stolyarov2023}, we also propose to couple the JPM to an additional dissipation channel---a long coplanar waveguide terminated by an absorbing element, \textit{e.g.}, normal metal~\cite{cattaneo2021,*lemz2024jltp}.
The waveguide is complemented by a side-coupled resonator whose frequency corresponds to the operating frequency of the JPM.
The resonator functions as a band-stop filter~\cite{reed2010}, suppressing the undesired relaxation of the lowest excited state in the shallow well, which is induced by the attached waveguide.
Such an add-on to the JPM design is essentially an \emph{engineered} bath with a reduced density of electromagnetic states near the operating frequency of the JPM.
Moreover, the proposed engineered bath functions as an ``exhaust pipe'', channeling the energy released during the JPM relaxation out of the system.
It prevents the JPM from heating up and mitigates the unwanted population of the resonator, to which the JPM is coupled, by the photons generated in the process of relaxation.
The modifications of the JPM design are shown in Fig.~\ref{fig:ModifiedJPM}.

\subsection{Model} \label{sec:JPMModel}
Let us now proceed to elaborate on the photomultiplier model.
We start by considering an uncoupled JPM.
We obtain the JPM eigenergies and eigenlevels by solving the stationary Schr\"{o}dinger equation.
Then, using the results for the uncoupled JPM, we address the case of a driven JPM capacitively coupled to a resonator mode.

\subsubsection{Uncoupled JPM} \label{sec:JPM_bare}
The rf-SQUID--based photomultiplier is described by the classical Hamiltonian
\begin{equation} \label{eq:jpm_clham}
	H_\pd = 4 E^\pd_{C} n^2_\pd + U_\pd,
\end{equation}
Here, we have introduced the reduced charge $n_\pd = Q_\pd/2e$, where $Q_\pd$ corresponds to a charge accumulated on a JPM capacitance.
The capacitive energy of the JPM is $E^\pd_{C} = e^2/2C_\pd$, where $C_\pd = C_\mathrm{S} + C^\pd_\JJ$ stands for its total capacitance.
The JPM potential energy reads as~\cite{opremcak2018, shnyrkov2018}
\begin{equation} \label{eq:U_jpm}
	U_\pd = \frac{1}{2} E^\pd_{L} \, \left(\phi_\pd -  2\pi\frac{\Phi_\mathrm{b}}{\Phi_0}\right)^2 - E^\pd_\JJ \cos \phi_\pd,
\end{equation}
where $E^\pd_{L} = L_\mathrm{S}^{-1} (\Phi_0/2\pi)^2$ is the JPM inductive energy and $\phi_\pd$ denotes the superconducting condensate phase drop over the JJ in the JPM.
The effect of the bias flux $\Phi_\mathrm{b}$ on the shape of the JPM potential is demonstrated in Fig.~\ref{fig:JPMSchemePotential}(b).

We determine the set of the JPM eigenstates $\{|\lambda\rangle\}$ and the corresponding eigenenergies $\{\cE_\lambda\}$ by numerically solving the stationary Schr\"{o}dinger equation:
\begin{equation} \label{eq:Schroedinger}
    \left[-\frac{1}{2}\frac{\partial^2}{\partial \phi_\pd^2} + u_\pd\right] \varPsi_\lambda = \varepsilon_\lambda \varPsi_\lambda,
\end{equation}
where $\varPsi_\lambda$ is the wavefunction of the eigenstate $|\lambda\rangle$.
Here, $u_\pd = U_\pd/8E^\pd_{C}$ and $\varepsilon_\lambda = \cE_\lambda/8E^\pd_{C}$ are the unitless JPM potential and the eigenenergy of the $\lambda$th eigenstate, respectively.

Next, we analyze the obtained JPM eigenstates to determine which eigenstates are localized in the left well, the right well, and above the potential barrier.
Here we assume the double-well asymmetric configuration of the JPM potential with the deeper right well.
The eigenstate $|\lambda\rangle$ belongs to a set of superbarrier states if its eigenenergy $\cE_\lambda$ exceeds the top of the potential barrier $\cE_\lambda > U_\pd(\phitop)$, otherwise this state is localized in one of the potential wells.
If the eigenenergy is lower than the bottom of the left potential well, $\cE_\lambda < U_\pd(\phibtm)$, such a state is localized in the right well.
Here, $\phitop$ denotes the phase ``coordinate'' of the top of the barrier, \textit{i.e.}, the local maximum of the JPM potential, while $\phibtm$ stands for the ``coordinate'' of the bottom of the left (shallow) well.
If $U_\pd(\phibtm) < \cE_\lambda < U_\pd(\phitop)$, then for this eigenstate we compute the average of the JPM phase variable $\langle\phi_{\mathrm{d}}\rangle_\lambda$.
For $\langle\phi_{\mathrm{d}}\rangle_\lambda < \phitop$, we conclude that such an eigenstate is localized in the left well, while in the opposite case, it is localized in the right well.
Figure~\ref{fig:JPMSchemePotential}(c) shows the region of the JPM asymmetric potential in the vicinity of the potential barrier.
It demonstrates the JPM eigenstates localized in the left (shallow) well, several upper eigenstates of the right (deep) well, and two lowest superbarrier eigenstates.

\subsubsection{Driven JPM coupled to resonator} \label{sec:JPM_res}
\begin{figure}[t!]
    \centering
    \includegraphics{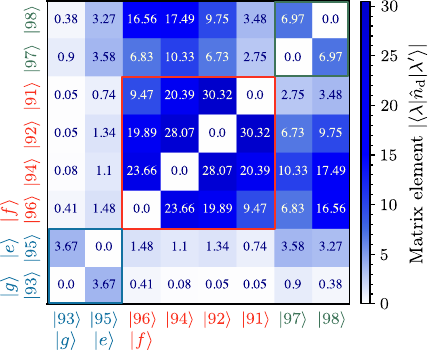}
    \caption{Magnitudes of the matrix elements $|\langle \lambda|\hat n_\pd|\lambda^\prime\rangle|$ of the reduced charge operator.
    The bias flux is $\Phib = 0.6316\Phi_0$.
    The rest of the JPM parameters are as in Table~\ref{tab:JPMParams}.}
    \label{fig:MatrixElements}
\end{figure}

Now, let us consider a driven JPM coupled to a mode of the buffer resonator via a shared capacitance $C_G$.
In this case, the JPM Hamiltonian reads as
\begin{equation} \label{eq:HamResJPM}
 \begin{split}
   \hat H_{\pd} = &\, \sum_\lambda \cE_\lambda \hat\sigma_{\lambda\lambda} + \hbar \sum_\lambda\sum_{\lambda^\prime\neq\lambda} \varOmega_{\lambda \lambda^\prime}\cos(\omega_\mathrm{dr} t) \hat\sigma_{\lambda^\prime\lambda} \\
   &\, + \hbar \sum_{\lambda}\sum_{\lambda^\prime\neq\lambda} G_{\lambda \lambda^\prime}(\hat a^\dag_2 - \hat a_2) \hat\sigma_{\lambda^\prime\lambda},
 \end{split}
\end{equation}
where we have introduced the operator $\hat \sigma_{\lambda^\prime\lambda} := |\lambda^\prime\rangle \langle \lambda|$.
The upper line of Eq.~\eqref{eq:HamResJPM} describes the JPM driven by the classical tone at frequency $\omega_\mathrm{dr}$, where $\varOmega_{\lambda \lambda'}$ is given by
\begin{equation} \label{eq:DriveAmplitude}
    \varOmega_{\lambda \lambda'} \approx 2\pi \langle\lambda|\hat n_\pd|\lambda^\prime\rangle \frac{C_x}{\tilde{C}_\pd} \frac{\mathcal{V}_\mathrm{dr}}{\Phi_0},
\end{equation}
where $\hat n_\pd = - \ii \partial/\partial \varphi_\pd$ is the JPM reduced charge operator.
Magnitudes of the matrix elements of the reduced charge operator for the JPM with parameters as in Fig.~\ref{fig:JPMSchemePotential} are presented in Fig.~\ref{fig:MatrixElements}.
Here, $\mathcal{V}_\mathrm{dr}$ stands for the amplitude of the drive voltage and $C_x$ is the coupling capacitance to the drive line as shown in Fig.~\ref{fig:ModifiedJPM}, while $\tilde{C}_\pd$ denotes the JPM loaded capacitance.
Assuming that the coupling capacitances are small compared to the JPM capacitance, $C_G, C_x, C_\kappa \ll C_\pd$, we approximate $\tilde{C}_\pd \approx C_\pd + C_G + C_x + C_\kappa$, where $C_\kappa$ is the coupling capacitance to the ``exhaust'' waveguide, cf. Sec.~\ref{sec:JPMModifications}.
Note that the JPM eigenenergies are now determined from Eq.~\eqref{eq:Schroedinger} with $\tilde{E}^\pd_C = e^2/2\tilde{C}_\pd$ instead of $E^\pd_C$.

The last term in Eq.~\eqref{eq:HamResJPM} describes the capacitive coupling between the JPM and the resonator.
Here, $G_{\lambda \lambda^\prime}$ stands for the coupling strength of $|\lambda\rangle \leftrightarrow |\lambda^\prime\rangle$ JPM transition to the resonator mode.
It is determined as
\begin{equation} \label{eq:Fll}
 G_{\lambda \lambda^\prime} = - \ii n_{\zpf 2} \langle \lambda|\hat n_\pd|\lambda^\prime\rangle \frac{E_g}{\hbar},
\end{equation}
where $E_g$ is the energy of the capacitive coupling between the JPM and the resonator.
Assuming that $C_G \ll \tilde{C}_2$, one has $E_g \approx 4e^2 C_G/(\tilde{C}^\prime_2 \tilde{C}_\pd)$, where $\tilde{C}^\prime_2 \approx \tilde{C}_2 + C_G$ stands for the capacitance of the buffer resonator renormalized by both the coupler self-capacitance (see Sec.~\ref{sec:TwoPhotonCoupling}) and the coupling capacitance to the JPM.

In our analysis, we focus on the transitions occurring between the pair of eigenstates localized in the left well and the uppermost eigenstate in the right well.
In addition, we are also interested in relaxation of these eigenestates to the lower eigenstates in the right well.
In what follows, for convenience, we relabel the eigenstates localized in the left well as $|g\rangle$ and $|e\rangle$ for the lower and upper states, respectively, while the upper eigenstate in the deep well is denoted as $|f\rangle$.
We assume that transitions to the superbarrier eigenstates are suppressed due to their strong detuning from both the resonator frequency and the drive frequency, as shown in Fig.~\ref{fig:LevelsRelaxation}(a).
Moreover, by deliberately enhancing the relaxation of the JPM eigenstates in the deep well, we increase the likelihood that excitation in the eigenstate $|f\rangle$ will ``roll'' down to lower eigenstates in the right well rather than jump to the superbarrier eigenstates.
Thus, for simplicity, in the Hamiltonian~\eqref{eq:HamResJPM} we neglect the terms involving the superbarrier eigenstates.
Besides that, we do not consider the population dynamics of the eigenstates in the deep well.\footnote{This problem is addressed in detail in Ref.~\cite{ilinskaya2024}.}
Hence, we also drop all the terms describing the transitions between the eigenstates localized in the right well except for the transitions from the uppermost eigenstate $|f\rangle$.
Such a reduced description significantly abates the numerical complexity of the problem while retaining the essential features of the JPM state evolution.
In the rotating-wave approximation (RWA), the reduced JPM Hamiltonian is given by
\begin{equation} \label{eq:HamJPM3LS}
  \begin{split}
    \hat{\cH}_\pd \approx & \, \hbar \omega_{ge} \hat\sigma_{ee} + \hbar \omega_{gf} \hat\sigma_{ff} - \hbar G (\hat a^\dag_2 \hat \sigma_{ge} + \hat\sigma_{eg}\hat a_2) \\
    & \, + \ii\hbar\Omega(\hat\sigma_{fe}\ee^{-\ii\omega_\mathrm{dr} t} - \hat\sigma_{ef}\ee^{\ii\omega_\mathrm{dr} t}),
   \end{split}
\end{equation}
where $\omega_{\lambda\lambda^\prime} = |\cE_{\lambda} - \cE_{\lambda^\prime}|/\hbar$ stands for the transition frequency between the eigenstates $|\lambda^\prime\rangle$ and $|\lambda\rangle$.
Hereafter, we also use the shorthand notations $G := |G_{ge}|$ for the coupling strength of the resonator mode to the $|g\rangle \leftrightarrow |e\rangle$ transition and $\Omega := |\varOmega_{ef}|/2$ for the drive strength.

\section{Master equation} \label{sec:MasterEquation}
To model the evolution of the entire system while accounting for incoherent processes, we employ the Lindblad master equation
\begin{equation} \label{eq:LindbladME}
 \frac{\partial \hat\rho}{\partial t} = \frac{\ii}{\hbar} \left[\hat\rho, \hat{\cH}_\mathrm{sys}\right] + \opD_\res\hat\rho + \opD_\pd \hat\rho.
\end{equation}
Here, $\hat{\cH}_\mathrm{sys}$ is the Hamiltonian of the coupled resonators-JPM system and the filter given by
\begin{equation} \label{eq:EffHamiltonian}
  \hat{\cH}_\mathrm{sys} = \hat{\cH}_{\res\res} + \hat{\cH}_{\pd} + \hat\cH_\mathrm{f},
\end{equation}
where the first term describing the dimer of inductively-coupled resonators is given by Eq.~\eqref{eq:hamrr_qnt4}, while the second term modeling the driven JPM coupled to the buffer resonator is expressed by Eq.~\eqref{eq:HamJPM3LS}.
Note that the frequency of the buffer resonator acquires an additional shift due to coupling to the JPM.
The last term in Eq.~\eqref{eq:EffHamiltonian} describes the filter resonator of frequency $\omega_\mathrm{f}$ and reads as $\hat\cH_\mathrm{f} = \hbar\omega_\mathrm{f} \hat f^\dag \hat f$, where $\hat f$ is the annihilation operator of a photon in this resonator.

\begin{figure*}[t!]
	\centering
	\includegraphics{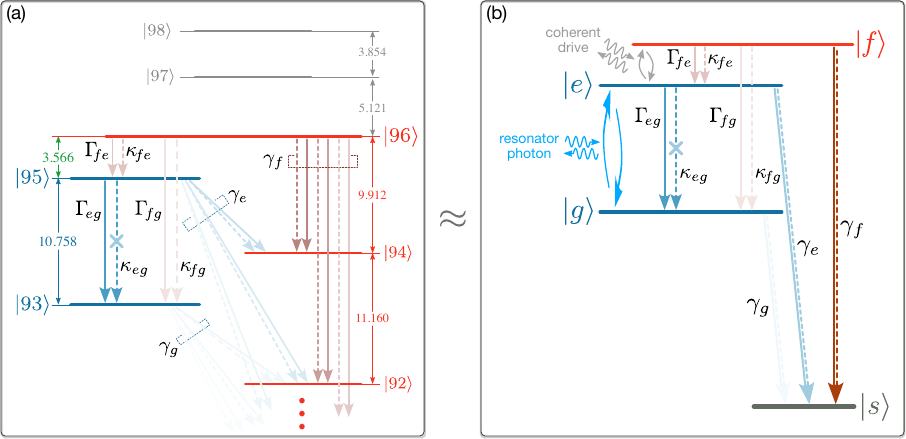}
	\caption{(a) Arrangement of the JPM eigenstates close to the potential barrier separating the wells and the relaxation paths of the relevant eigenstates $|g\rangle$, $|e\rangle$, and $|f\rangle$.
        Solid (dashed) arrows show spontaneous relaxation paths induced by the internal (external) bath.
        Relaxation of the eigenstate $|e\rangle$ to the eigenstate $|g\rangle$ induced by the external bath is inhibited due to the Purcell filter.
        Higher saturation of arrows indicates higher rate of a corresponding irreversible transition.
        The transition frequencies are given in units of $2\pi\,\mathrm{GHz}$.
        The JPM parameters are the same as in Fig.~\ref{fig:JPMSchemePotential}.
        (b) Simplified model of the detector:
        Coupling to the mode of the buffer resonator entails coherent transitions between the eigenstates $|g\rangle$ and $|e\rangle$, while the coherent drive induces transitions between the eigenstates $|e\rangle$ and $|f\rangle$.
        Multiple relaxation paths of the eigenstates $|g\rangle$, $|e\rangle$, and $|f\rangle$ to the lower eigenstates localized in the deep well shown in Fig.~\ref{fig:LevelsRelaxation}(a) are replaced with the relaxation to the auxiliary fictitious state $|s\rangle$ with rates $\gamma_g$, $\gamma_e$, and $\gamma_f$, respectively.
        \label{fig:LevelsRelaxation}}
\end{figure*}

In the master equation~\eqref{eq:LindbladME}, the terms $\opD_\res \hat\rho$ and $\opD_\pd \hat \rho$ describe the incoherent processes in the resonators dimer and the JPM, respectively.
The superoperator $\opD_\res$ describing the losses in the resonators is given by\footnote{Hereinafter, we assume that the number of thermal excitations in the system is negligible.
Indeed, for the typical working temperatures $T_\mathrm{s} \approx 10-30\,\mathrm{mK}$ and operational frequencies $\omega_\mathrm{s}/2\pi \approx 5-10\,\mathrm{GHz}$ of the circuit QED setups, the number of thermal excitations in the system is $n_\mathrm{th} < 10^{-2}$.}
\begin{equation} \label{eq:DissipatorRes}
    \opD_\res = \sum_{j=1}^2 \varGamma_j \mathsf{D}[\hat a_j],
\end{equation}
where $\mathsf{D}[\hat\bullet]\hat\rho := \hat\bullet^\dag\hat\rho\hat\bullet - (\hat\bullet^\dag\hat\bullet\hat\rho + \hat\rho\hat\bullet^\dag\hat\bullet)/2$ denotes the standard Lindbladian superoperator \cite{breuer2002book, manzano2020}.
We denote the single-photon loss rate of the $j$th resonator as $\varGamma_j$.

Superoperator $\opD_\pd$ attributed to the JPM is written as $\opD_\pd = \opD^\downarrow_\pd + \opD^\phi_\pd$, where the first term describes the relaxation of the JPM eigenstates, while the second term models their \emph{pure} dephasing.
The dissipation superoperator $\opD^\downarrow_\pd$ accounts for the contributions from both the JPM internal dissipative bath and the external engineered bath.
It reads as (see Appendix~\ref{sec:JPMDissipator} for the detailed derivation)
\begin{equation} \label{eq:DissipatorJPMRelax}
  \begin{split}
    \opD^\downarrow_\pd = & \, \Gamma_{e g}\mathsf{D}[\hat\sigma_{ge}] + \mathsf{D}[\sqrt{\kappa_{e g}}\hat\sigma_{ge} + \sqrt{\kappa_\mathrm{f}} \hat f] \\
    & \, + (\Gamma_{f e} + \kappa_{f e})\mathsf{D}[\hat\sigma_{ef}] + (\Gamma_{f g} + \kappa_{f g})\mathsf{D}[\hat\sigma_{gf}] \\
    & \, + \gamma_{g} \mathsf{D}[\hat\sigma_{s g}] + \gamma_{e} \mathsf{D}[\hat\sigma_{s e}] + \gamma_{f} \mathsf{D}[\hat\sigma_{s f}].
  \end{split}
\end{equation}
The terms in the upper line describe the relaxation of the eigenstate $|e\rangle$ to the eigenstate $|g\rangle$.
The first term corresponds to relaxation due to the internal losses of the JPM, while the second term describes relaxation to the engineered bath---``exhaust'' waveguide equipped with the band stop filter.
Here, $\Gamma_{\lambda \lambda^\prime}$ and $\kappa_{\lambda \lambda^\prime}$ denote the relaxation rates of the eigenstate $|\lambda\rangle$ to the eigenstate $|\lambda^\prime\rangle$ due to internal and engineered losses, respectively, while $\kappa_\mathrm{f}$ stands for the filter linewidth.
In our analysis, we assume that the JPM internal bath is Ohmic.
Thus, the internal relaxation rates are related as~\cite{you2007, jones2013}
\begin{equation}
  \frac{\Gamma_{\lambda\lambda^\prime}}{\Gamma_{\mu\mu^\prime}} = \frac{\omega_{\lambda\lambda^\prime}}{\omega_{\mu\mu^\prime}} \left|\frac{\langle\lambda|\hat n_\pd|\lambda^\prime\rangle}{\langle \mu|\hat n_\pd|\mu^\prime\rangle}\right|^2,
\end{equation}
where $\cE_{\lambda^\prime} < \cE_\lambda$ and $\cE_{\mu^\prime} < \cE_{\mu}$.
For our model (see Appendix~\ref{sec:JPMDissipator}), the relaxation rates due to coupling to the waveguide obey the similar relationship as the internal relaxation rates, \textit{i.e.}, $\kappa_{\lambda\lambda^\prime}/\kappa_{\mu\mu^\prime} = \Gamma_{\lambda\lambda^\prime}/\Gamma_{\mu\mu^\prime}$.
The second line in Eq.~\eqref{eq:DissipatorJPMRelax} describes relaxation of the eigenstate $|f\rangle$ to the eigenstates $|e\rangle$ and $|g\rangle$.
The bottom line in Eq.~\eqref{eq:DissipatorJPMRelax} describes relaxation of the eigenstates $|g\rangle$, $|e\rangle$, and $|f\rangle$ to the lower eigenstates localized in the right well, where $\gamma_{\lambda} = \tilde{\Gamma}_{\lambda} + \tilde{\kappa}_\lambda$ is the \emph{total} relaxation rate of the eigenstate $|\lambda\rangle$ ($\lambda\in\{g,e,f\}$) to the lower eigenstates localized in the right well, with $\tilde{\Gamma}_\lambda = \sum_{\lambda^\prime} \Gamma_{\lambda\lambda^\prime}$ and $\tilde{\kappa}_\lambda = \sum_{\lambda^\prime} \kappa_{\lambda\lambda^\prime}$.
Here, the sum goes over the eigenstates localized in the right well whose eigenenergy is lower than that of the eigenstate $|\lambda\rangle$.
Table~\ref{tab:RelaxationRates} demonstrates the relaxation rates of the eigenstates $|g\rangle$, $|e\rangle$, and $|f\rangle$ due to internal (in units of $\Gamma_{eg}$) and engineered (in units of $\kappa_{eg}$) losses.

Following Ref.~\cite{stolyarov2023}, we introduce the auxiliary fictitious state $|s\rangle$ in the dissipator~\eqref{eq:DissipatorJPMRelax} to simplify the representation of the cascade of eigenstates in the right well [see Fig.~\ref{fig:LevelsRelaxation}(a)]. This substitution accounts for the relaxation pathways of the eigenstates $|g\rangle$, $|e\rangle$, and $|f\rangle$.
Similar approach was used in Refs.~\cite{govia2012, schoendorf2018} and \cite{sokolov2020} to describe CBJJ-based JPMs.
Such an approach significantly shrinks the Hilbert space of JPM states, reducing the numerical complexity of the problem while retaining the relevant features of the JPM state evolution.
This allows us to model the JPM as a four-level system as shown in Fig.~\ref{fig:LevelsRelaxation}(b).

\begin{table}[t]
	\caption{Internal relaxation rates of the relevant JPM eigenstates in units of the relaxation rate $\Gamma_{eg}$ of the eigenstate $|e\rangle$ to the eigenstate $|g\rangle$.
        The JPM parameters are the same as those used in Figs.~\ref{fig:JPMSchemePotential}(c) and~\ref{fig:MatrixElements}.
        \label{tab:RelaxationRates}}
	\begin{center}
		\begin{ruledtabular}
            \newcolumntype{4}{D{.}{.}{4}} 
                \renewcommand{\arraystretch}{1.1}
			\begin{tabular}{ccccc}
				    $\Gamma_{fg}/\Gamma_{eg}$ & $\Gamma_{fe}/\Gamma_{eg}$ & $\tilde{\Gamma}_{f}/\Gamma_{eg}$ & $\tilde{\Gamma}_{e}/\Gamma_{eg}$ & $\tilde{\Gamma}_{g}/\Gamma_{eg}$ \\
                    $\kappa_{fg}/\kappa_{eg}$ & $\kappa_{fe}/\kappa_{eg}$ & $\tilde{\kappa}_{f}/\kappa_{eg}$ & $\tilde{\kappa}_{e}/\kappa_{eg}$ & $\tilde{\kappa}_{g}/\kappa_{eg}$ \\
                    \hline
                    0.0184 & 0.0458 & 121.56 & 0.4817 & 0.0007 \\
			\end{tabular}
		\end{ruledtabular}
	\end{center}
\end{table}

We assume that the dominant mechanism of pure dephasing is the Gaussian $1/f$ noise in the external flux through the JPM loop~\cite{ithier2005,yoshihara2006,bialczak2007}.
In this case, superoperator $\opD^\phi_\pd$ describing the JPM pure dephasing reads as \cite{didier2019, dipaolo2021, groszkowski2023}
\begin{equation} \label{eq:DissipatorJPMDeph}
    \opD^\phi_\pd = 2\varGamma^2_{e} t \, \mathsf{D}[\hat\sigma_{e e}] + 2\varGamma^2_{f} t \, \mathsf{D}[\hat\sigma_{f f}],
\end{equation}
where $\varGamma_\lambda$ denotes pure dephasing rate of eigenstate $|\lambda\rangle$ with $\lambda\in\{e,f\}$.
In Eq.~\eqref{eq:DissipatorJPMDeph}, linear time dependence ensures the Gaussian decay law of the non-diagonal matrix elements (coherences) of the density operator.
Such a time dependence is characteristic for the Gaussian $1/f$ noise~\cite{martinis2003, ithier2005, omalley2015, anton2012, sete2017, danilin2024}.
The dephasing rates are estimated as $\varGamma_\lambda = \sqrt{\zeta_\lambda} \cA_{\Phi} \left|\partial \omega_{\lambda g}/\partial \Phib\right|$, \textit{cf.} Refs.~\cite{martinis2003, ithier2005, anton2012, omalley2015, sete2017}.
Here, $\cA_\Phi$ is the amplitude of the flux noise, typically ranging from $10^{-6} \Phi_0$ to $10^{-5} \Phi_0$~\cite{rkoch2007, hutchings2017, kou2017}.
Numerical factor $\zeta_\lambda$ is determined as $\zeta_\lambda \approx \ln(2.516\varGamma_\lambda/\varpi_\mathrm{cut})$, where $\varpi_\mathrm{cut}$ stands for the low-frequency cutoff of the noise spectrum, usually set at $\varpi_\mathrm{cut}/2\pi = 1\,\mathrm{Hz}$~\cite{martinis2003, omalley2015, anton2012, sete2017}.
For $\Phib = 0.6316\Phi_0$ and the JPM parameters provided in Table~\ref{tab:JPMParams}, we estimate the pure dephasing rates of eigenstates $|e\rangle$ and $|f\rangle$ to be $\varGamma_e/2\pi \approx 1.3\,\textrm{MHz}$ and $\varGamma_f/2\pi \approx 30\,\textrm{MHz}$, respectively, for the noise amplitude $\cA_\Phi = 10^{-6} \Phi_0$.

\begin{figure*}[t!]
	\centering
	\includegraphics{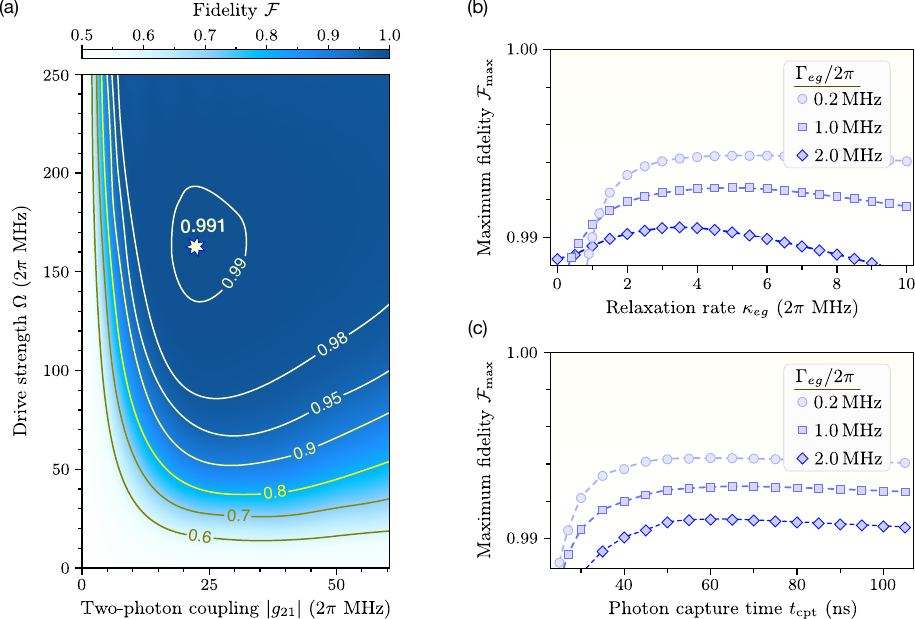}
	\caption{Effect of system settings on detection fidelity.
        (a) Dependence of the detection fidelity $\mathcal{F}$ on the two-photon coupling strength $|g_{21}|$ and drive strength $\Omega$.
        For calculations, here we set the rates $\Gamma_{eg}/2\pi = 1\,\mathrm{MHz}$ and $\kappa_{eg}/2\pi = 1\,\mathrm{MHz}$, while the photon capture time is $t_\mathrm{cpt} = 50\,\mathrm{ns}$.
        Star marks the maximum fidelity $\mathcal{F}_\mathrm{max}$.
        (b) Effect of the JPM relaxation on $\mathcal{F}_\mathrm{max}$ for $t_\mathrm{cpt} = 50\, \mathrm{ns}$.
        (c) Dependence of $\mathcal{F}_\mathrm{max}$ on the photon capture time $t_\mathrm{cpt}$ for $\kappa_{eg}/2\pi = 3\,\mathrm{MHz}$.
        The remaining setup parameters used for calculations are discussed in Sec.~\ref{sec:Performance}.
        \label{fig:FidelityResults}}
\end{figure*}

\section{Photodetector performance} \label{sec:Performance} 

To characterize the ability of the considered detector to distinguish a two-photon state from single-photon and vacuum states, we use the detection fidelity $\mathcal{F}$ which corresponds to the probability of correct discrimination of a two-photon state.
Specifically, it should click when there are at least two photons in the storage resonator and should not click when there are fewer than two photons.
This probability is given by
\begin{equation} \label{eq:Fidelity0}
  \mathcal{F}  = p_2 P_{\clk|2} + p_1 (1 - P_{\clk|1}) + p_0 (1 - P_{\clk|0}),
\end{equation}
where $p_n$ is the probability of finding $n$ photons in the input state, \textit{i.e.}, the initial (at $t=0$) state of the storage resonator, and $P_{\clk|n}$ denotes the JPM click probability for the $n$-photon Fock-state input.
We assume that the probabilities of the JPM click for the vacuum and single-photon inputs are equal, $(P_{\clk|1} = P_{\clk|0}) \equiv P_{\clk|<2}$.
This follows from the fact that in both cases the buffer resonator will be in the vacuum state, and the non-zero probability of a false click arises from the relaxation of the state $|g\rangle$ localized in the shallow well of the JPM potential to the states localized in the deep well.
This assumption allows us to rewrite Eq.~\eqref{eq:Fidelity0} as
\begin{equation} \label{eq:Fidelity1}
    \mathcal{F} = p_2 P_{\clk|2} + p_{<2} (1-P_{\clk|<2}),
\end{equation}
where $p_{<2} = p_0 + p_1$ is the probability that initially there are less than two photons in the storage resonator.
Since we have no prior information about the measured state, we assume that there is an equal probability that the resonator is prepared in a two-photon state and that the resonator is prepared in a state with less than two photons, \textit{i.e.}, $p_2 = p_{<2} = 1/2$.
Under this assumption, Eq.~\eqref{eq:Fidelity1} acquires the form
\begin{equation} \label{eq:Fidelity}
    \mathcal{F} = \frac{1 + P_{\clk|2} - P_{\clk|<2}}{2}.
\end{equation}
This definition of fidelity is analogous to that used for quantifying the efficiency of qubit readout (see, \textit{e.g.}, Ref.~\cite{hazra2025arxiv} and references therein).

The probability the JPM has clicked after photon capture time $t_\mathrm{cpt}$ for the $n$-photon (where $n\geq 2$) input state is determined as~\cite{stolyarov2023}
\begin{equation} \label{eq:Pclk_n}
	P_{\clk|n\geq 2} = \eta \, \Tr\left[\hat\rho_{|n\rangle}(t_\mathrm{cpt})\hat\sigma_{ss}\right],
\end{equation}
where $\eta \in [0..1]$ is the phenomenological parameter modeling the efficiency of the JPM flux state readout.
By $\hat\rho_{|n\rangle}(t)$, we denote the system density operator at time $t$ for the $n$-photon input state.
For determining the density operator $\hat\rho_{|n\rangle}(t)$, we solve the Lindblad master equation~\eqref{eq:LindbladME} numerically using \textsf{QuantumOptics.jl}~\cite{kramer2018quantumoptics}---the versatile quantum optics toolkit for \textsc{Julia}~\cite{bezanson2017julia}.
For simulations, we assume that initially the state of the system is separable with the two-photon Fock state of the storage resonator, the vacuum state of the buffer resonator, and the JPM residing in its eigenstate $\left|g\right\rangle$, \textit{i.e.}, $\hat\rho_{|2\rangle}(0) = \left|2\right\rangle_{1}\!\left\langle 2\right| \otimes \left|0\right\rangle_{2}\!\left\langle 0\right| \otimes \left|g\right\rangle\!\left\langle g\right|$, where $\left|n\right\rangle_j$ denotes the $n$-photon Fock state of the $j$th resonator.

If the input state contains less than two photons, the buffer resonator remains in the vacuum state, and the JPM can deliver a click only due to relaxation of the eigenstate $|g\rangle$ to the lower eigenstates localized in the deep well.
The Lindblad master equation~\eqref{eq:LindbladME} with the dissipator~\eqref{eq:DissipatorJPMRelax} suggests that in this case, the (false) click probability is given by~\cite{sokolov2020, stolyarov2023}
\begin{equation}\label{eq:FalseCountProbability}
    P_{\clk|<2} = \eta \, (1 - \ee^{-\gamma_g t_\mathrm{cpt}}).
\end{equation}
The expression in parentheses corresponds to the probability of \emph{spontaneous} relaxation of the eigenstate $|g\rangle$ to the fictitious state $|s\rangle$ over the photon capture time $t_\mathrm{cpt}$.

Figure~\ref{fig:FidelityResults} aggregates the results that demonstrate the dependence of the detection fidelity on the characteristics of the setup.
Figure~\ref{fig:FidelityResults}(a) demonstrates the dependence on the coupler and drive parameters.
Computations reveal that for given values of the photon capture time $t_\mathrm{cpt}$, the JPM-resonator coupling strength $G$, and the relaxation rates $\Gamma_{eg}$ and $\kappa_{eg}$, there exists a combination of the JPM drive strength $\Omega$ and the two-photon coupling strength $|g_{21}|$ that maximizes the detection fidelity.
This dependence of the fidelity on the setup parameters reflects the complex interplay of the key processes in the system under consideration.
Let us elucidate this result qualitatively.
On the one hand, too weak two-photon coupling between the resonators would yield slow conversion of a photon pair from the storage resonator into a single photon in the buffer resonator, making the reference capture time insufficient.
On the other hand, making this coupling too strong would cause the photon in the buffer resonator more likely to be converted back into a pair of photons in the storage resonator rather than excite the JPM, which slows down detection.
For a drive that is too weak, excitation in the eigenstate $|e\rangle$ is more likely to decay to the eigenstate $|g\rangle$, emitting a photon back into the buffer resonator, rather than transitioning to $|f\rangle$ and subsequently cascading down to the lowest eigenstates in the right well, which would produce a click.
Conversely, with excessively strong driving, excitation in the eigenstate $|f\rangle$ is more likely to transition back to the eigenstate $|e\rangle$ rather than begin relaxing to the lower eigenstates in the right well.
Both of these scenarios deteriorate detection fidelity.

Figure~\ref{fig:FidelityResults}(b) illustrates the effect of an external engineered bath on the maximum detection fidelity.
In general, the introduction of engineered decay enhances the detector's efficiency as intended [see Sec.~\ref{sec:JPMModifications}].
However, if the relaxation rate becomes too large, the false count rate [see Eq.~\eqref{eq:FalseCountProbability}] overpowers the positive effect.
The results also indicate that when the decay rate to the waveguide is slow, increasing the JPM internal losses improves detection fidelity by accelerating relaxation to the eigenstates in the right well.
Conversely, for faster engineered decay, higher internal losses lead to reduced fidelity.

The dependence of the maximum fidelity on the photon capture time is shown in Fig.~\ref{fig:FidelityResults}(c).
The results indicate that an initial increase of the photon capture time $t_\mathrm{cpt}$ rapidly enhances maximum fidelity.
However, beyond a certain point, the fidelity gradually declines as $t_\mathrm{cpt}$ continues to increase.
This behavior arises because, at this stage, a further increase of $t_\mathrm{cpt}$ no longer affects the probability of a correct click $P_{\clk|2}$, while it increases the probability of an erroneous click $P_{\clk|<2}$, as described by Eq.~\eqref{eq:FalseCountProbability}.
Consequently, the detection fidelity decreases, as follows from Eq.~\eqref{eq:Fidelity}.

\begin{table}[t]
	\caption{Sample sets of the simulation parameters that ensure detection fidelity exceeding $99\%$ for the photon capture time up to 50 ns.
        \label{tab:SetupParameters}}
	\begin{center}
		\begin{ruledtabular}
            \newcolumntype{4}{D{.}{.}{4}} 
                \renewcommand{\arraystretch}{1.1}
			\begin{tabular}{lcc}
				    Parameter & \multicolumn{2}{c}{Value} \\
				    \hline
				    & Set $A$ & Set $B$ \\
                    \cline{2-3}
                    %
                    %
                    $G/2\pi$ & 50.0 MHz & 60.0 MHz \\
                    $g_{21}/2\pi$ & 20.4 MHz & 24.4 MHz \\
                    \hline
                    %
                    %
                    $\omega_1/2\pi$ & \multicolumn{2}{c}{5.379 GHz} \\
                    $K_1/2\pi$ & 277.4 kHz & 396.9 kHz \\
                    $\varGamma_1/2\pi$ & 10.0 kHz & 2.0 kHz \\
                    %
                    \hline
					$\omega_2/2\pi$ & \multicolumn{2}{c}{10.758 GHz} \\
                    $K_2/2\pi$ & 138.7 kHz & 198.5 kHz \\
                    $\varGamma_2/2\pi$ & 100.0 kHz & 20.0 kHz \\
                    \hline
                    %
                    $\omega_{ge}/2\pi$ & \multicolumn{2}{c}{10.758 GHz} \\
                    $\omega_\mathrm{dr}/2\pi$ & \multicolumn{2}{c}{3.566 GHz} \\
                    $\Omega/2\pi$ & 220.6 MHz & 188.2 MHz \\
                    $\Gamma_{eg}/2\pi$ & 1.0 MHz & 0.1 MHz \\
                    $\kappa_{eg}/2\pi$ & 4.0 MHz & 5.0 MHz \\
                    $\eta$ & 99.5\% & 99.9\% \\
                    $t_\mathrm{cpt}$ & 50 ns & 30 ns \\
                    \hline
                    $\mathcal{F}$ & 99.24\% & 99.79\%
			\end{tabular}
		\end{ruledtabular}
	\end{center}
\end{table}

For simulations in Fig.~\ref{fig:FidelityResults}, we assume the resonant regime of operation $2\omega_1 = \omega_2 = \omega_{ge}$.
For the resonator-JPM coupling strength, we choose a reasonable value of $G/2\pi = 50\,\mathrm{MHz}$.
In all calculations, we set $\kappa_\mathrm{f} = 100\kappa_{eg}$, which ensures effective suppression of unwanted relaxation of the eigenstate $|e\rangle$ to the eigenstate $|g\rangle$ due to emission into the waveguide. 
The efficiency of the JPM flux state readout is set $\eta = 99.5\%$, which is consistent with Refs.~\cite{opremcak2018, opremcak2021} reporting readout efficiency over $99.9\%$.
We set the single-photon loss rate of the storage resonator to be $\varGamma_1/2\pi = 10\,\mathrm{kHz}$ corresponding to the internal quality factor of about $5\times10^5$ for the frequency near $2\pi \times 5\,\mathrm{GHz}$.
Lumped-element resonators with even higher quality factors were demonstrated in experiment (see, \textit{e.g.}, Refs.~\cite{lilishi2022, crowley2023, dhundhwal2025arxiv}).
For the buffer resonator, we set higher loss rate $\varGamma_2/2\pi = 100\,\mathrm{kHz}$ to account for the effect of tunable elements discussed in Sec.~\ref{sec:Setup}.
For the resonator frequency near $2\pi \times 10\,\mathrm{GHz}$, this corresponds to the internal quality factor of about $10^5$.

Furthermore, we assume the internal relaxation rate of the JPM excited eigenstate $|e\rangle$ to be $\Gamma_{eg}/2\pi \leq 2\,\mathrm{MHz}$, which corresponds to the internal $Q$-factor $>5 \times 10^3$ for the JPM operating frequency we consider.
Note that in the experiment~\cite{opremcak2018}, the loss rate \( \Gamma_{eg} \approx 2\pi \times 15\,\mathrm{MHz} \) for the flux-biased JPM is significantly higher than the values used in our calculations.
Such high internal losses were necessary therein to ensure rapid relaxation to the bottom of the right well after photon absorption.
On the contrary, for the considered design with an engineered bath, it is beneficial to use a high-$Q$ JPM for achieving high detection fidelity.
We anticipate that with different materials such as tantalum~\cite{place2021, wang2022, wang2025prappl} and refined fabrication techniques~\cite{wang2022, bal2024}, it is possible to considerably reduce the JPM internal losses and achieve the values we used in the simulations.

In the choice of simulation parameters, we are constrained by the validity criteria of the RWA used in deriving the effective Hamiltonian of the system in Eq.~\eqref{eq:EffHamiltonian}.
Thus, the strength of the two-photon coupling should be small compared to the resonator frequencies, \textit{i.e.}, $|g_{21}| \ll \omega_1, \omega_2$, while the JPM-resonator coupling should satisfy $G \ll \omega_2, \omega_{ge}$.
The drive tone should be strongly off-resonant for the $|g\rangle \leftrightarrow |e\rangle$ transition, therefore, its strength should be much weaker than the detuning $\Delta = \omega_{ge} - \omega_{ef}$ between the frequencies of $|g\rangle \leftrightarrow |e\rangle$ and $|e\rangle \leftrightarrow |f\rangle$ transitions, \textit{i.e.}, $|\varOmega_{eg}| \ll \Delta$.
On the other hand, the drive strength for the $|e\rangle \leftrightarrow |f\rangle$ transition should be weak compared to the transition frequency, $\Omega \ll \omega_{ef}$.
Moreover, coupling between the mode of the buffer resonator and the target $|e\rangle \leftrightarrow |f\rangle$ transition should be strongly off-resonant as well with $|G_{ef}| \ll \Delta$.

To sum up the quantitative analysis of the performance of the considered detection scheme, we present Table~\ref{tab:SetupParameters}, demonstrating two sets ($A$ and $B$) of the simulation parameters with photon capture time up to 50 ns, for which the detection fidelity exceeds 99$\%$.
Among the two, set $A$ constitutes a more moderate selection of parameters based on those used for obtaining data for Fig.~\ref{fig:FidelityResults}.
Compared to set $A$, set $B$ represents a more optimistic estimate with better JPM readout efficiency, five times higher resonator quality factors, and an order of magnitude improvement in the JPM internal loss rate.
In addition, we also choose a slightly stronger coupling between the JPM and the buffer resonator.
These improvements allow one to expect higher fidelity for an even shorter photon capture time compared to set $A$.
In both sets, the values of the two-photon coupling and drive strengths correspond to those that maximize fidelity.

\section{Summary and Outlook} \label{sec:summ}
To summarize, we have theoretically demonstrated a feasible circuit QED design of a microwave two-photon threshold detector.
The proposed scheme effectively transforms a flux-biased JPM, operating as a single-photon threshold detector, into a detector with two-photon threshold.
Our design incorporates an auxiliary resonant two-photon conversion process mediated by a nonlinear inductive coupler.
Additionally, an appropriate choice of the coupler flux bias enables the suppression of superfluous linear inductive and cross-Kerr couplings between the resonators.
To improve the detector performance, we introduced several modifications to the JPM design.
First, we reduced the false count rate by deepening the left well.
Second, we applied a resonant drive tone to the transition between the upper eigenstates in the left and right wells to maintain a high photon absorption rate. Furthermore, to accelerate JPM relaxation to its global energy minimum and mitigate setup heating, we proposed coupling the JPM to an engineered bath---a waveguide with a band-stop filter.
The filter effectively suppresses emission into the waveguide at the JPM operating frequency.

We derived the master equation governing the dynamics of the model system. 
As a figure of merit, characterizing the efficiency of the considered scheme, we used detection fidelity---the probability of correct discrimination of the two-photon input state from the single-photon and vacuum states.
By analyzing the system's state evolution, we determined the click probability which allowed us to calculate the detection fidelity and study its dependence on the system parameters.
The results of calculations suggest that detection fidelity exceeding 99$\%$ can be achieved for setup parameters accessible for current circuit QED technologies.

Although we put forward a rather specific design of a two-photon detector, which is built around the SPD represented by a flux-biased JPM, we emphasize that the considered scheme is compatible with other types of SPDs that support strong capacitive coupling to a resonator mode.
In particular, instead of the flux-biased JPM design, one can use a CBJJ-based JPM.
In addition, transmon-based SPDs \cite{besse2018, dass2020prappl, lesc2020, balembois2024}, ultrasensitive microwave bolometers \cite{kokkoniemi2020, gunyho2024} and calorimeters \cite{pekola2022} can be utilized as well.
Moreover, a promising design of a microwave SPD utilizing criticality-enhanced sensing in the Kerr Josephson parametric amplifier was experimentally demonstrated in Ref.~\cite{petrovnin2024}.

Let us now briefly outline an extension of the considered detection scheme.
In Ref.~\cite{stolyarov2023}, an approach for achieving limited photon-number resolution with just one JPM was considered.
It relies on repeated measurements of a resonator mode by a flux-biased JPM.
To speed up counting and improve photon-number resolution, one can use several JPMs, or other SPDs, simultaneously probing the same mode.
In this case, one can resolve up to $M N_1$ photons in $M$ measurement iterations, where $N_1$ is the number of JPMs (SPDs).
This arrangement is a circuit QED analog of an array (or click) detector (see, \textit{e.g.}, Refs.~\cite{sperling2012, kovtoniuk2024} and references therein) widely used in quantum optics.
Elaborating this idea even further, one can construct a click detector consisting of one JPM operating as an SPD and one or several JPMs working as two-photon detectors.
Such a hybrid detector allows one to count resonator photons up to $M (2N_2 + 1)$ in $M$ measurement cycles, where $N_2$ is a number of JPMs operating in the two-photon regime.
To achieve the same resolution with this detector, one would require $N_\mathrm{2}$ less JPMs than for the click detector composed solely of the JPMs operating as SPDs, which relieves the hardware requirements for implementing a microwave photon number resolving detector.

The concept of transforming an SPD into a multiphoton threshold detector can be extended even further.
Several proposals have aimed at isolating specific multiphoton interactions as dominant by cascading lower-order nonlinear processes or applying a coherent drive~\cite{vanselow2025arxiv, Mundhada2017, *Mundhada2019}.
By utilizing coherent multiphoton conversion processes beyond the proposed two-photon interactions, it is, in principle, possible to design a multiphoton threshold detector.
In this scenario, a click detector composed of threshold detectors for different photon numbers would offer improved photon-number resolution while reducing the number of required measurements.

\begin{acknowledgments}
 The authors thank A.~Miano and D.~Weiss for careful reading of the manuscript and useful comments.
 E.~V.~S. thanks A.~Sokolov for numerous fruitful discussions on various aspects of circuit QED, A.~Semenov for enlightening discussions on photodetection, and O.~Kliushnichenko for helpful suggestions on numerical methods.
 R.~A.~B. thanks S.~Girvin and M.~Devoret for insightful discussions and helpful comments on the results.
 E.~V.~S. acknowledges support from the National Research Foundation of Ukraine through the Project No. 2023.03/0165, Quantum correlations of electromagnetic radiation.
 R.~A.~B. acknowledges support from the Army Research Office (ARO) under Grant Number W911NF-23-1-0051, from the Air Force Office of Scientific Research (AFOSR) under grant no. FA9550-19-1-0399, and from the U.S. Department of Energy (DoE), Office of Science, National Quantum Information Science Research Centers, Co-design Center for Quantum Advantage (C2QA) under contract number DE-SC0012704.
 The views and conclusions contained in this document are those of the authors and should not be interpreted as representing the official policies, either expressed or implied, of the ARO, AFOSR, DoE or the US Government.
 The US Government is authorized to reproduce and distribute reprints for Government purposes notwithstanding any copyright notation herein.
\end{acknowledgments}


\appendix

\section{Derivation of the Hamiltonian for resonators coupled via BiSQUID} \label{sec:HamiltonianDerivation}

\begin{figure}[t!] 
	\centering
	\includegraphics{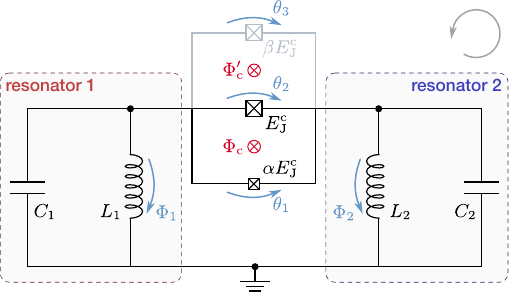}
	\caption{Circuit diagram of two resonators coupled via the BiSQUID.
        Arrows indicate the direction of the flux drop.
        All loops in the circuit are propagated counterclockwise. \label{fig:CircuitScheme}}
\end{figure}

Here, we provide derivation of the quantum Hamiltonian in Eq.~\eqref{eq:QuantumHamResRes} describing a circuit comprised by two single-mode resonators inductively coupled via the BiSQUID (or asymmetric dc SQUID).
The diagram of this circuit is demonstrated in Fig.~\ref{fig:CircuitScheme}.
We start by writing down the Lagrangian of the circuit $\mathcal{L}_{\res\res}$, which is given by \cite{devoret1997, vool2017, rasmussen2021}
\begin{equation} \label{eq:Lagrangian}
	\begin{split}
	  \mathcal{L}_{\res\res} = & \, \sum_{j=1}^2 \left(\frac{C_j \dot\Phi_j^2}{2} - \frac{\Phi_j^2}{2L_j}\right) \\
        & \, + \frac{C^\mathrm{c}_{\alpha\mathrm{J}} \dot \theta_1^2}{2} + \alpha E^\mathrm{c}_\JJ \cos\left(2\pi\frac{\theta_1}{\Phi_0}\right) \\
        & \, + \frac{C^\mathrm{c}_\JJ\dot\theta_2^2}{2}
        + \frac{E^\mathrm{c}_\JJ}{1+\beta} \cos\left(2\pi\frac{\theta_2}{\Phi_0}\right) \\
        & \, + \frac{\beta C^\mathrm{c}_\JJ \dot\theta_3^2}{2} + \frac{\beta E^\mathrm{c}_\JJ}{1+\beta}\cos\left(2\pi\frac{\theta_3}{\Phi_0}\right),
	\end{split}
\end{equation}
where the first term corresponds to the Lagrangian of the resonators with $\Phi_j$ being the flux drop over the $j$th resonator capacitance ($j\in\{1,2\}$).
The next two terms describe the ``black sheep" junction of the BiSQUID, where $\theta_1$ is the flux drop across this JJ with $C^\cpl_{\alpha\JJ}$ being its self-capacitance.
The last two lines in Eq.~\eqref{eq:Lagrangian} describe, depending on the value of $\beta$, either a larger junction of the asymmetric SQUID or a flux-biased symmetric SQUID loop of the BiSQUID.
Here, we have introduced the binary parameter $\beta \in \{0,1\}$ to set the coupler type we are using: the asymmetric SQUID ($\beta = 0$) or the BiSQUID ($\beta=1$).

Using the fluxoid quantization condition \cite{tinkham2004,krantz2019,rasmussen2021}, one obtains $\theta_1 = \Phi_1 - \Phi_2$, $\theta_2 = \theta_1 - \Phi_\cpl$, and $\theta_3 = \theta_2 - \Phi^\prime_\cpl$.
Substituting these relationships in Eq.~\eqref{eq:Lagrangian} yields
\begin{equation} \label{eq:Lagrangian2}
  \begin{split}
    \mathcal{L}_{\res\res} = & \, \sum_{j=1}^2 \left[\frac{(C_j + C_\cpl) \dot\Phi_j^2}{2} - \frac{\Phi_j^2}{2L_j}\right] \\ & \, - C_\cpl \dot{\Phi}_1 \dot{\Phi}_2 + E^{\cpl, \beta}_{\JJ\,\eff} \cos\left(2\pi \frac{\Phi_1 - \Phi_2}{\Phi_0} - \vartheta_\beta\right),
  \end{split}
\end{equation}
where $C_\cpl = C^\cpl_{\alpha\JJ} + (1+\beta)C^\cpl_{\JJ}$ denotes the total capacitance of the coupler.
We consider the external fluxes that thread the coupler to be constant.
In practice, however, we have to turn the fluxes on and off.
The assumption of constant fluxes implies that they are switched adiabatically, \textit{i.e.}, on timescales much longer than the inverse of the lowest characteristic frequency in the system, which, in our case, is the frequency of the storage resonator.
This requirement ensures that excitation of spurious photons in the resonators is avoided.
Given the GHz operating frequencies of the circuit QED systems, the external fluxes can be switched within a few nanoseconds.
The effective Josephson energy of the coupler $E^{\cpl, \beta}_{\JJ\,\eff}$ and the phase shift $\vartheta_\beta$ are given by Eqs.~\eqref{eq:EJBiSQUID} and \eqref{eq:PhaseShiftBiSQUID}, where $\Phi^\prime_\cpl$ is replaced with $\beta \Phi^\prime_\cpl$.

Performing the Legendre transform \cite{goldstein1980}, one obtains the classical Hamiltonian of the circuit:
\begin{equation} \label{eq:legnd}
	H_{\res\res} = \sum_{j=1}^2 Q_j \dot\Phi_j  - \mathcal{L}_{\res\res},
\end{equation}
where $Q_j = \partial \mathcal{L}_{\res\res}/\partial \dot\Phi_j$ denotes the generalized momentum corresponding in this case to the charge accumulated on the $j$th resonator capacitance.
These charges are determined as
\begin{equation} \label{eq:Qs}
	\left(\!
	\begin{array}{l}
		Q_1 \\
		Q_2
	\end{array}
	\!\right)
	=
	\begin{pmatrix}
		C_1 + C_\cpl && -C_\cpl \\
		-C_\cpl && C_2 + C_\cpl
	\end{pmatrix}
	\left(\!
	\begin{array}{l}
		\dot{\Phi}_1 \\
		\dot{\Phi}_2
	\end{array}
	\!\right).
\end{equation}
Substituting Eqs.~\eqref{eq:Lagrangian2} into \eqref{eq:legnd} along with expressing voltage drops $\dot{\Phi}_j$ in terms of $Q_j$ using Eq.~\eqref{eq:Qs}, and then introducing the reduced charge and flux variables $n_j = Q_j/(2e)$ and $\phi_j = 2\pi \Phi_j/\Phi_0$, one arrives at the circuit Hamiltonian given by
\begin{equation} \label{eq:ClassicalHamResRes}
	\begin{split}
		H_{\res\res} = & \, \sum_{j=1}^2 \left(4 \tilde{E}_{Cj} n^2_j + \frac{1}{2} E_{Lj} \phi^2_j\right) \\
		& \, + E^\cpl_{C} n_1 n_2 - E^\mathrm{c}_{\JJ\,\mathrm{eff}} \cos (\phi_1 - \phi_2 - \vartheta_\beta).
	\end{split}
\end{equation}
The expression in parentheses represents the total electromagnetic energy of the $j$th resonator.
We have introduced the unitless charge and flux variables $n_j = Q_j/2e$ and $\phi_j = 2\pi \Phi_j/\Phi_0$.
Here, $Q_j$ is the charge accumulated on the capacitance of the $j$th resonator, while $\Phi_j$ is the flux drop across the corresponding resonator related to the voltage drop $V_j$ on its capacitance as $\dot\Phi_j = V_j$ \cite{devoret1997, vool2017, rasmussen2021}.
Thus, $n_j$ is the number of Cooper pairs on the $j$th resonator capacitance, while $\phi_j$ is interpreted as the gauge-invariant condensate phase drop across this resonator.

In Eq.~\eqref{eq:ClassicalHamResRes}, $\tilde{E}_{Cj} = e^2/2 \tilde{C}_j$ and $E_{Lj} = L^{-1}_j (\Phi_0/2\pi)^2$ are the renormalized capacitive and bare inductive energy of the $j$th resonator, respectively.
Here, $L_j$ is the bare inductance of the $j$th resonator, and $\tilde{C}_j$ denotes its loaded capacitance determined as
\begin{equation} \label{eq:C1C2}
    \tilde{C}_1 = C_1 + \frac{C_2 C_\cpl}{C_2 + C_\cpl}, \quad \tilde{C}_2 = C_2 + \frac{C_1 C_\cpl}{C_1 + C_\cpl},
\end{equation}
where $C_j$ stands for the $j$th resonator bare capacitance, and $C_\cpl = C^\cpl_\JJ + C^\cpl_{\alpha \JJ}$ is the total capacitance of the coupler.
Typically, the coupler capacitance is small compared to the resonator capacitances, $C_\cpl \ll C_1, C_2$, which yields an approximation $\tilde{C}_j \approx C_j + C_\cpl$.

The terms in the second line of Eq.~\eqref{eq:ClassicalHamResRes} describe the capacitive and inductive couplings between the resonators, mediated by the asymmetric SQUID coupler.
The former is attributed to the capacitances of the coupler's JJs with $E^\cpl_{C}$ being the capacitive coupling energy given by
\begin{equation} \label{eq:ecap}
	E^\cpl_{C} = \frac{4e^2}{C_1 C_2 (C_1^{-1} + C_2^{-1} + C_\cpl^{-1})} \approx 4 e^2 \frac{C_\cpl}{C_1 C_2},
\end{equation}
where the approximation holds in the regime of small coupler capacitance, as realized in the system under consideration.
For the latter, $E^\mathrm{c}_{\JJ\,\mathrm{eff}}$ is the effective Josephson energy of the coupler given by Eq.~\eqref{eq:EJ_eff}.
The coupler-specific phase shift $\vartheta_\beta$ appearing in the last term of the Hamiltonian~\eqref{eq:ClassicalHamResRes} is expressed by Eq.~\eqref{eq:PhaseShift} for the case of the asymmetric SQUID coupler ($\beta = 0$) and by Eq.~\eqref{eq:PhaseShiftBiSQUID} for the BiSQUID coupler ($\beta = 1$).
Note that in the main text, we drop the subscript $\beta$ for brevity.

\begin{figure}[t!] 
	\centering
	\includegraphics{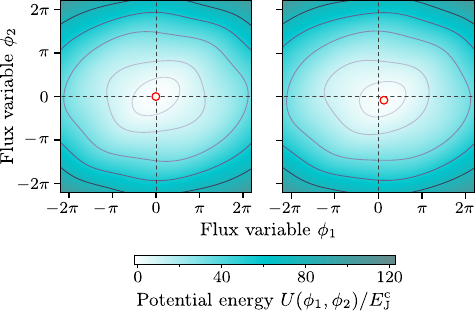}
	\caption{Potential energy of the resonators dimer for (left plot) $\Phi_\cpl = 0$ and (right plot) $\Phi_\cpl = \Phi_0/4$.
		Circles indicate the potential energy minima.
		Lines of equal energy are shown in solid color.
		Parameters of the circuit are as follows: $\alpha = 0.4$, $E_{L1}/E^\cpl_\JJ = 1.5$, and $E_{L1}/E_{L2} = 2$.
		\label{fig:CircuitPotential}}
\end{figure}

The inductive terms in Eq.~\eqref{eq:ClassicalHamResRes} constitute the potential energy $U(\phi_1, \phi_2)$ of the circuit \cite{devoret1997, vool2017}:
\begin{equation} \label{eq:Urr}
	U(\phi_1, \phi_2) = \frac{1}{2}\sum_{j=1}^2 E_{L j} \phi^2_j - E^\cpl_{\JJ\,\mathrm{eff}} \cos(\phi_1 - \phi_2 - \vartheta_\beta).
\end{equation}
Typically, the inductive energies of the resonators dominate their capacitive energies $E_{Lj} \gg E_{Cj}$.
The mechanical analogy for this regime is a heavy particle trapped in a deep potential well, oscillating near its equilibrium point, which corresponds to the local minimum of the potential energy $U(\phi_1, \phi_2)$.
In our further analysis, we consider the circuit parameters for which the inductive energies of the resonators exceed the Josephson coupling energy $E_{Lj} > E^\cpl_{\JJ\,\eff}$.
In this case, the potential energy of the circuit $U(\phi_1, \phi_2)$ has a single minimum, whose position $(\phi^\mathrm{min}_1, \phi^\mathrm{min}_2)$ is found by solving a set of transcendental equations:
\begin{equation} \label{eq:kirch}
	\begin{split}
		& E_{L1} \phi^\mathrm{min}_1 - E^\mathrm{c}_{\JJ\,\mathrm{eff}} \sin(\phi^\mathrm{min}_1 - \phi^\mathrm{min}_2 - \vartheta_\beta) = 0, \\
		& E_{L2} \phi^\mathrm{min}_2 + E^\mathrm{c}_{\JJ\,\mathrm{eff}} \sin(\phi^\mathrm{min}_1 - \phi^\mathrm{min}_2 - \vartheta_\beta) = 0.
	\end{split}
\end{equation}
Up to a factor of $2\pi/\Phi_0$, the above equations manifest the Kirchhoff's current law \cite{feynman_lectures, rasmussen2021}.
For the case when the external flux equals an integer or half-integer multiple of the flux quantum, \textit{i.e.}, $\Phi_\cpl/\Phi_0 = \mathsf{n}/2$ with $\mathsf{n} \in \mathbb{Z}$, Eq.~\eqref{eq:kirch} yields $\phi^\mathrm{min}_{1,2} = 0$, which coincides with the case of the inductively uncoupled resonators $E^\cpl_\JJ = 0$.
However, when the inductive coupling is on, $E^\cpl_\JJ > 0$, and $\Phi_\cpl/\Phi_0 \neq  \mathsf{n}/2$, the position of the potential minimum shifts compared to that for the uncoupled regime, as illustrated in Fig.~\ref{fig:CircuitPotential}.

Following the standard quantization approach for the electrical circuits \cite{devoret1997, vool2017, rasmussen2021, blais2021}, we proceed to the quantum Hamiltonian of the system by promoting charge $n_j$ and flux $\phi_j$ variables to the non-commuting quantum operators obeying the commutation relation $[\hat \phi_j, \hat n_l] = \ii \delta_{j,l}$, where $j, l \in \{1, 2\}$.
Next, it is convenient to formally represent the flux operator as $\hat\phi_j = \phi^\mathrm{min}_j + \hat\varphi_j$ with $j \in \{1,2\}$.
Using this representation in the quantized version of the system Hamiltonian~\eqref{eq:ClassicalHamResRes}, and expanding the cosine term into a Taylor series around the potential minimum, while taking into account Eq.~\eqref{eq:kirch}, we obtain the Hamiltonian
\begin{equation} \label{eq:QuantumHamResRes}
 \begin{split}
    \hat{H}_{\res\res} = & \, \sum_{j=1}^2 \left(4 \tilde{E}_{Cj} \hat n^2_j + \frac{1}{2} \tilde{E}_{Lj} \hat \varphi^2_j\right) + E^\cpl_C \, \hat n_1 \hat n_2 \\ & \, + E^\mathrm{c}_{\JJ\,\mathrm{eff}} u_2 \hat \varphi_1 \hat \varphi_2
    - E^\mathrm{c}_{\JJ\,\mathrm{eff}} \sum_{k \geq 3} \frac{u_k}{k!} (\hat\varphi_1 - \hat\varphi_2)^k.
 \end{split}
\end{equation}
Here, we have rearranged the terms and introduced the notation $\tilde{E}_{Lj} = E_{Lj} - E^\mathrm{c}_{\JJ\,\mathrm{eff}} u_2$ for the renormalized inductive energy of the $j$th resonator.

To proceed, we express the charge and the flux operators of the resonators via the bosonic ladder operators as $\hat n_j = - \ii n_{\zpf j}(\hat a^\dag_j - \hat a_j)$ and $\hat\varphi_j = \varphi_{\zpf j}(\hat a^\dag_j + \hat a_j)$.
Plugging the definitions of the operators $\hat n_j$ and $\hat \varphi_j$ in terms of the ladder operators into the Hamiltonian~\eqref{eq:QuantumHamResRes} and rearranging the terms, one obtains
Eq.~\eqref{eq:HamiltonianResRes}.

\section{Derivation of the JPM dissipator} \label{sec:JPMDissipator}
In this appendix, we provide the Hamiltonian that models the JPM dissipation.
This Hamiltonian is then used to derive the Lindbladian dissipator given by Eq.~\eqref{eq:DissipatorJPMRelax}.

\subsection{JPM bath Hamiltonian} \label{sec:HamJPMBath}

In this appendix, we discuss the model of the JPM dissipation.
We model this process by coupling the JPM to two \emph{uncorrelated} bosonic baths, where one is responsible for the internal JPM losses, while the other represents the effect of the engineered electromagnetic environment coupled to the JPM, as described in Sec.~\ref{sec:JPMModifications}.
The Hamiltonian describing these baths and their interactions with the JPM reads as
\begin{equation} \label{eq:BathHamiltonian}
  \hat{H}_\mathrm{b} = \hat{H}_\mathrm{ib} + \hat{H}_\mathrm{eb}.
\end{equation}
In the RWA, the Hamiltonian of the internal dissipative bath is given by~\cite{breuer2002book}
\begin{equation} \label{eq:HamInternalBath}
 \begin{split}
\hat{H}_\mathrm{ib} = & \, \hbar \int\limits_0^\infty \dd\nu \nu \hat b^\dag_{\nu} \hat b_{\nu} \\
 & \, + \hbar \int\limits_0^\infty \dd\nu \sum_{\lambda\in\boldsymbol{\Xi}}\sum_{\lambda^\prime<\lambda} B_{\lambda\lambda^\prime}(\nu) (\hat b^\dag_{\nu}\hat \sigma_{\lambda^\prime\lambda} + \hat\sigma_{\lambda\lambda^\prime}\hat b_{\nu}),
 \end{split}
\end{equation}
where for the sake of brevity, we have introduced a notation $\boldsymbol{\Xi} = \{g,e,f\}$, while the notation $\textstyle \sum_{\lambda^\prime<\lambda}$ implies that sum goes over the eigenstates $\lambda^\prime$ satisfying $\cE_{\lambda^\prime} < \cE_{\lambda}.$
The first term models the bath of independent bosonic modes with $\hat b_\nu$ being the annihilation operator of an excitation in the mode with frequency $\nu$.
The second term describes the coupling between the bath and the JPM, where $B_{\lambda\lambda^\prime}(\nu)$ denotes the coupling strength of the $|\lambda\rangle\leftrightarrow|\lambda^\prime\rangle$ transition to the bath mode with frequency $\nu$.
For the Ohmic bath and assuming that $k_\mathrm{B} T_\mathrm{s} \ll \hbar\nu$, one has $B_{\lambda\lambda^\prime}(\nu) \propto \sqrt{\nu}|\langle \lambda|\hat n_\pd|\lambda^\prime\rangle|$ \cite{blais2021}.

The Hamiltonian for the external engineered bath is expressed as
\begin{equation} \label{eq:EngineeredBath}
 \begin{split}
 \hat{H}_\mathrm{eb} = & \, \hbar \int\limits_0^\infty \dd\nu \nu \hat w^\dag_{\nu} \hat w_{\nu}
 + \hbar \int\limits_0^\infty \dd\nu F(\nu) (\hat w^\dag_\nu \hat f + \hat f^\dag \hat w_\nu) \\
  & \, + \hbar \int\limits_0^\infty \dd\nu \sum_{\lambda\in\boldsymbol{\Xi}}\sum_{\lambda^\prime<\lambda} W_{\lambda\lambda^\prime}(\nu) (\hat w^\dag_{\nu} \hat\sigma_{\lambda^\prime\lambda} + \hat\sigma_{\lambda\lambda^\prime} \hat w_\nu),
  \end{split}
\end{equation}
where the upper line describes the ``exhaust" waveguide coupled to the filter resonator.
The operator $\hat w_\nu$ annihilates an excitation in the waveguide mode of frequency $\nu$, and $F(\nu)$ stands for the coupling strength between the filter resonator and the waveguide mode.
The bottom line describes the capacitive coupling of the waveguide to the JPM, where $W_{\lambda\lambda^\prime}(\nu)$ is the strength of capacitive coupling between the waveguide mode of frequency $\nu$ and the transition $|\lambda^\prime\rangle \leftrightarrow |\lambda\rangle$.
Far from the waveguide cutoff frequency \cite{parrarodriguez2018, shitara2021}, the JPM-waveguide coupling is well approximated by \cite{blais2021}
\begin{equation} \label{eq:WaveguideCoupling}
 \begin{split}
    W_{\lambda\lambda^\prime}(\nu) \approx &\, |\langle \lambda|\hat n_\pd|\lambda^\prime\rangle| \sqrt{\frac{Z_\mathrm{w}}{Z_\pd}} \frac{C_\kappa}{\tilde{C}_\pd} \sqrt{\frac{\nu}{2\pi}}, 
 \end{split}
\end{equation}
where $Z_\mathrm{w}$ and $Z_\pd$ are the waveguide and the JPM impedance, respectively, while $C_\kappa$ is the coupling capacitance of the JPM to the waveguide and $\tilde{C}_\pd$ is the JPM loaded capacitance.

\subsection{Lindblad master equation} \label{sec:LindbladME}

As a first step in the derivation of the master equation, we consider the incoherent dynamics between the JPM and the external engineered bath described by the Hamiltonian
\begin{equation} \label{eq:JPMBathHamiltonian}
\hat H_{\mathrm{db}} = \sum_{\lambda\in\boldsymbol{\Xi}} \cE_\lambda \hat\sigma_{\lambda\lambda} + \hat{H}_\mathrm{eb},
\end{equation}
which can be split into the diagonal part
\begin{equation}
  \hat{H}_0 = \sum_{\lambda\in\boldsymbol{\Xi}} \cE_\lambda \hat\sigma_{\lambda\lambda}+\hbar\omega_\mathrm{f} \hat f^\dag \hat f
  + \hbar \int\limits_0^\infty \dd\nu \nu \hat w^\dag_{\nu} \hat w_{\nu}
\end{equation}
containing a bare Hamiltonian for the JPM, the filter resonator, and the ``exhaust'' waveguide, and the interaction part given by
\begin{equation} \label{eq:H_int_bath}
 \begin{split}
  \hat{H}_\mathrm{I} = & \,  \hbar \int\limits_0^\infty \dd\nu F(\nu) (\hat w^\dag_\nu \hat f + \hat f^\dag \hat w_\nu) \\ & \, + \hbar \int\limits_0^\infty \dd\nu \sum_{\lambda\in\boldsymbol{\Xi}}\sum_{\lambda^\prime<\lambda} W_{\lambda\lambda^\prime}(\nu) (\hat w^\dag_{\nu}\hat\sigma_{\lambda^\prime\lambda} + \hat\sigma_{\lambda\lambda^\prime} \hat w_{\nu}).
  \end{split}
\end{equation}

To obtain the master equation, it is useful to switch to the interaction picture by applying the time-dependent unitary transformation $\hat{U}_{I}^\dag\hat H_{\mathrm{db}}\hat{U}_{I}-\ii\hat{U}_{I}^\dag\dot{\hat{U}}_{I}\rightarrow\hat H_\mathrm{I}(t)$ with $\hat{U}_\mathrm{I} = \exp{(\ii \hat{H}_0 t)}$, where the transformed Hamiltonian is given by
\begin{equation}\label{eq:ham_int_time}
  \hat{H}_\mathrm{I}(t) = \hbar \int\limits_0^\infty \dd\nu \sum_{\lambda\in\boldsymbol{\Xi}}\sum_{\lambda^\prime<\lambda} \left[\ee^{\ii(\nu-\omega_{\lambda\lambda^\prime})t} \hat w^\dag_\nu \hat\varSigma_{\lambda^\prime \lambda}(\nu) + \mathhc\right].
\end{equation}
For the sake of brevity, we introduce the operators
\begin{equation}
\hat \varSigma_{\lambda\lambda^\prime}(\nu) = 
 \left\{
 \begin{array}{ll}
   W_{\lambda\lambda^\prime}(\nu)\hat\sigma_{\lambda\lambda^\prime}, \qquad \quad \, \textrm{for}\,|g\rangle\nleftrightarrow|e\rangle \\
   W_{ge}(\nu)\hat\sigma_{ge} + F(\nu)\hat f, \, \textrm{for}\, |g\rangle\leftrightarrow|e\rangle
 \end{array}
 \right.
\end{equation}
where we accounted for the effect of the band-stop filter, which interferes with the $|g\rangle \leftrightarrow |e\rangle$ transition due to the proximity of the frequencies $\omega_{ge} \approx \omega_\mathrm{f}$.

Next,  we define the density operator of the JPM-waveguide-filter system as $\hat{\varrho}_
\mathrm{dwf}(t) = \hat{\varrho}_\mathrm{df}(t) \otimes \hat{\varrho}_\mathrm{w}(0)$, where $\hat{\varrho}_\mathrm{df}$ represents the joint density operator for the JPM and the filter resonator, and $\hat{\varrho}_\mathrm{w}$ denotes the waveguide density operator, assuming Born approximation, \textit{i.e.}, the waveguide eigenstates remain unchanged due to relatively weak coupling to the JPM and the filter~\cite{Zhang2016,manzano2020}.
In this case, the Markovian master equation for the density operator of the JPM-filter subsystem in the interaction picture is given by \cite{manzano2020}
\begin{equation} \label{markovian_me}
  \frac{\partial \hat\varrho_\mathrm{df}(t)}{\partial t} = -\frac{1}{\hbar^2}\int\limits_{0}^\infty \dd \tau \, \Tr_\mathrm{w}\left[\hat{H}_\mathrm{I}(t),\left[\hat{H}_\mathrm{I}(t-\tau),\hat{\varrho}_
\mathrm{dwf}(t)\right]\right],
\end{equation}
where we utilize the standard assumption that $\hat\varrho_\mathrm{df}(t)$ does not depend on the past history of the density operator and apply the trace over the waveguide operators.

Substituting $\hat{H}_\mathrm{I}(t)$ from Eq.~\eqref{eq:ham_int_time} into Eq.~\eqref{markovian_me}, we rewrite the master equation as \cite{Brasil2013}
\begin{widetext}
\begin{equation}
\begin{split}
 \frac{\partial \hat\varrho_\mathrm{df}}{\partial t} = & \sum_{\lambda\in\boldsymbol{\Xi}}  \sum_{\lambda^\prime<\lambda} \sum_{\mu\in\boldsymbol{\Xi}} \sum_{\mu^\prime<\mu} \int\limits_{0}^\infty \dd\nu \int\limits_{0}^\infty \dd \tau \\
 & \, \times \left[\ee^{\ii(\nu-\omega_{\lambda\lambda^\prime})\tau -\ii(\omega_{\mu\mu^\prime}-\omega_{\lambda\lambda^\prime})t} \Tr_\mathrm{w}(\hat{w}^\dag_\nu\hat{w}_\nu\hat{\varrho}_\mathrm{w}) \left(\hat \varSigma_{\mu\mu^\prime}(\nu)\hat{\varrho}_\mathrm{df}\hat \varSigma_{\lambda\lambda^\prime}^\dag(\nu) - \hat \varSigma_{\mu\mu^\prime}(\nu)\hat \varSigma_{\lambda\lambda^\prime}^\dag(\nu)\hat{\varrho}_\mathrm{df}\right)\right. \\ 
 & \qquad \left. + \, \ee^{\ii(\nu-\omega_{\mu\mu^\prime})\tau + \ii(\omega_{\mu\mu^\prime}-\omega_{\lambda\lambda^\prime})t} \Tr_\mathrm{w}(\hat{w}_\nu\hat{w}^\dag_\nu\hat{\varrho}_\mathrm{w}) \left(\hat \varSigma_{\mu\mu^\prime}^\dag(\nu)\hat{\varrho}_\mathrm{df}\hat \varSigma_{\lambda\lambda^\prime}(\nu)-\hat \varSigma_{\mu\mu^\prime}^\dag(\nu)\hat \varSigma_{\lambda\lambda^\prime}(\nu)\hat{\varrho}_\mathrm{df}\right) + \mathhc\right].
\end{split}
\end{equation}
For deriving the above expression, we used that for a thermal bath $\Tr_\mathrm{w}(\hat{w}^\dag_\nu\hat{w}_{\nu^\prime}\hat\rho_\mathrm{w})=\Tr_\mathrm{w}(\hat{w}^\dag_\nu\hat{w}_{\nu}\hat\rho_\mathrm{w})\delta(\nu-\nu^\prime)$ and performed the respective integration over $\nu^\prime$.
To further simplify the expression, we perform the remaining integration and introduce the notation ${n}_\mathrm{th}(\nu):=\Tr_\mathrm{w}(\hat{w}^\dag_\nu\hat{w}_\nu\hat{\rho}_\mathrm{w})$ corresponding to the average number of thermal photons in the waveguide mode of frequency $\nu>0$ \cite{Clerk2010}.
This leads us to the following result
\begin{align}\label{b7}
 & \frac{\partial \hat\rho_\mathrm{df}}{\partial t}= \pi \sum_{\lambda\in\boldsymbol{\Xi}}  \sum_{\lambda^\prime<\lambda} \sum_{\mu\in\boldsymbol{\Xi}} \sum_{\mu^\prime<\mu} \ee^{\ii(\omega_{\lambda\lambda^\prime} - \omega_{\mu\mu^\prime})t} 
 {n}_\mathrm{th}(\omega_{\lambda\lambda^\prime}) \left(2\hat\varSigma_{\mu\mu^\prime}(\omega_{\lambda\lambda^\prime}) \hat{\rho}_\mathrm{df}\hat \varSigma_{\lambda\lambda^\prime}^\dag(\omega_{\lambda\lambda^\prime}) -
 \left\{\hat \varSigma_{\mu\mu^\prime}(\omega_{\lambda\lambda^\prime})\hat \varSigma_{\lambda\lambda^\prime}^\dag(\omega_{\lambda\lambda^\prime}),\hat{\rho}_\mathrm{df}\right\}\right) \nonumber
 \\ 
 & \quad + \pi \sum_{\lambda\in\boldsymbol{\Xi}} \sum_{\lambda^\prime<\lambda} \sum_{\mu\in\boldsymbol{\Xi}} \sum_{\mu^\prime<\mu} \ee^{\ii(\omega_{\mu\mu^\prime}-\omega_{\lambda\lambda^\prime})t} ({n}_\mathrm{th}(\omega_{\mu\mu^\prime})+1)  \left(2\hat \varSigma_{\mu\mu^\prime}^\dag(\omega_{\mu\mu^\prime})\hat{\rho}_\mathrm{df}\hat \varSigma_{\lambda\lambda^\prime}(\omega_{\mu\mu^\prime})
 - \left\{\hat\varSigma_{\mu\mu^\prime}^\dag(\omega_{\mu\mu^\prime}) \hat\varSigma_{\lambda\lambda^\prime}(\omega_{\mu\mu^\prime}), \hat{\rho}_\mathrm{df}\right\}\right),
\end{align}
where $\{\hat A, \hat B\} = \hat A \hat B + \hat B \hat A$ denotes the anticommutator. 
\end{widetext}
For the derivation, we employed an expression
\[\int\limits_{0}^\infty \dd \tau \,\ee^{\pm\ii(\nu-\omega_{\lambda\lambda^\prime})\tau} = \pi\delta(\nu-\omega_{\lambda\lambda^\prime}) \pm \frac{\ii}{(\nu-\omega_{\lambda\lambda^\prime})}\]
and omitted the second term.
This term contributes to the bath-induced Lamb shift, which can be neglected since it is much smaller than the JPM transition frequencies \cite{Brasil2013, McCauley2020}.

If the system lacks processes with closely spaced transition energies, then, in the spirit of the RWA approximation, the dynamics will be dominated by processes where $\omega_{\mu\mu^\prime} = \omega_{\lambda\lambda^\prime}$.
Fast oscillating terms with $|\omega_{\mu\mu^\prime} - \omega_{\lambda\lambda^\prime}| \gg |W_{\mu\mu^\prime}(\omega_{\lambda\lambda^\prime})W_{\lambda\lambda^\prime}^*(\omega_{\lambda\lambda^\prime})|$, meaning they vary much faster than the characteristic timescales of system evolution, can therefore be neglected \cite{manzano2020}.
Then, the expression in Eq.~\eqref{b7} reduces to
\begin{equation} \label{eq:rwa_lindbladian}
 \begin{split}
    \frac{\partial \hat\varrho_\mathrm{df}}{\partial t} = &\, \sum_{\lambda\in\boldsymbol{\Xi}}\sum_{\lambda^\prime<\lambda}^{|g\rangle\nleftrightarrow|e\rangle}\kappa_{\lambda\lambda^\prime}({n}_\mathrm{th}(\omega_{\lambda\lambda^\prime}){+}1)\mathsf{D}[\sigma_{\lambda\lambda^\prime}]\hat{\varrho}_\mathrm{df} \\
    &\, + \sum_{\lambda\in\boldsymbol{\Xi}}\sum_{\lambda^\prime<\lambda}^{|g\rangle\nleftrightarrow|e\rangle}\kappa_{\lambda\lambda^\prime}{n}_\mathrm{th}(\omega_{\lambda\lambda^\prime})\mathsf{D}[\sigma_{\lambda^\prime\lambda}]\hat{\varrho}_\mathrm{df} \\
    &\, + ({n}_\mathrm{th}(\omega_{ge}){+}1)\mathsf{D}[\sqrt{\kappa_{eg}}\sigma_{ge}{+}\sqrt{\kappa_\mathrm{f}}\hat f]\hat{\varrho}_\mathrm{df} \\
    &\, + {n}_\mathrm{th}(\omega_{ge})\mathsf{D}[\sqrt{\kappa_{eg}}\sigma_{eg}{+}\sqrt{\kappa_\mathrm{f}}\hat f^\dag]\hat{\varrho}_\mathrm{df},
 \end{split}
\end{equation}
where the decay rates for the JPM and the filter resonator are $\kappa_{\lambda\lambda^\prime}=2\pi|W_{\lambda\lambda^\prime}(\omega_{\lambda\lambda^\prime})|^2$ and $\kappa_\mathrm{f}=2\pi|F(\omega_{ge})|^2$, respectively. The first two terms represent incoherent processes in the JPM attributed to the engineered bath, excluding the $|g\rangle \leftrightarrow |e\rangle$ transition, which is described separately by terms that account for interference with the band-stop filter.

Analogous to the effect of the external bath on the JPM, the same approach applies to the incoherent dynamics induced by the internal bath, yielding the decay rates $\Gamma_{\lambda\lambda^\prime} = 2\pi|B_{\lambda\lambda^\prime}(\omega_{\lambda\lambda^\prime})|^2$ [see Eq.~\eqref{eq:HamInternalBath}] for the corresponding transitions.
Since two baths are uncorrelated, we can write down the dissipation superoperator for the joint density operator of the JPM and the filter
\begin{equation} \label{eq:DissipatorJPMfull}
  \begin{split}
    \opD^\downarrow_\pd = & \, \Gamma_{e g}\mathsf{D}[\hat\sigma_{ge}] + \mathsf{D}[\sqrt{\kappa_{e g}}\hat\sigma_{ge} + \sqrt{\kappa_\mathrm{f}} \hat f] \\
     & \, + \sum_{\lambda\in\boldsymbol{\Xi}}\sum_{\lambda^\prime<\lambda}^{|g\rangle\nleftrightarrow|e\rangle}(\Gamma_{\lambda\lambda^\prime} + \kappa_{\lambda\lambda^\prime})\mathsf{D}[\sigma_{\lambda\lambda^\prime}],
  \end{split}
\end{equation}
where $n_\mathrm{th}(\omega_{\lambda\lambda^\prime})$ is assumed to be zero.

Focusing on the dynamics of the $|g\rangle$, $|e\rangle$, and $|f\rangle$ eigenstates of the JPM, where decay to any other eigenstate localized in the right well corresponds to a ``click'' event, the superoperator in Eq.~\eqref{eq:DissipatorJPMfull} can be simplified to Eq.~\eqref{eq:DissipatorJPMRelax}. This reduction is achieved by introducing an auxiliary fictitious state $|s\rangle$ (see Sec.~\ref{sec:MasterEquation}) that accounts for the relaxation of the eigenstate $\lambda\in\{g,e,f\}$ to the eigenstates localized in the right well with an effective rate $\gamma_{\lambda} = \sum_{\lambda^\prime} (\Gamma_{\lambda\lambda^\prime} + \kappa_{\lambda\lambda^\prime})$.


\bibliography{bibliography}

\end{document}